\documentclass[prd,reprint,nofootinbib,notitlepage,aps,tightenlines,preprintnumbers,amsmath,amssymb,showpacs,superscriptaddress]{revtex4-1}
\usepackage{graphicx,epstopdf,amsmath,amsfonts,amssymb,multirow}
\usepackage[usenames,dvipsnames]{color}

\newcommand{\nc}{\newcommand}
\nc{\AF}[1]{\textcolor{blue}{[AF: #1]}}
\nc{\MP}[1]{\textcolor{red}{[MP: #1]}}
\nc{\AR}[1]{\textcolor{cyan}{[AR: #1]}}
\nc{\JP}[1]{\textcolor{blue}{[JP: #1]}}
\nc{\beq}{\begin{equation}}
\nc{\eeq}{\end{equation}}

\nc{\alphad}{\alpha_{\rm d}}
\nc{\alphap}{\alpha'}
\nc{\neff}{N_{\rm eff}}
\nc{\dneff}{\Delta \neff}
\nc{\Mpl}{M_{\rm pl}}
\nc{\mA}{m_{A'}}

\nc{\nvec}[1]{\mathrm{#1}}

\def\ga{\mathrel{\raise.3ex\hbox{$>$\kern-.75em\lower1ex\hbox{$\sim$}}}}
\def\la{\mathrel{\raise.3ex\hbox{$<$\kern-.75em\lower1ex\hbox{$\sim$}}}}
\def\gappeq{\mathrel{\rlap {\raise.5ex\hbox{$>$}}
{\lower.5ex\hbox{$\sim$}}}}
\def\lappeq{\mathrel{\rlap{\raise.5ex\hbox{$<$}}E_f E_\bar{f} - \sqrt{E_f^2 - m_f^2}\sqrt{E_\bar{f}^2 - m_f^2}
{\lower.5ex\hbox{$\sim$}}}}

\def\gyr{{\rm \, G\kern-0.125em yr}}

\def\MeV{{\rm \, MeV}}
\def\keV{{\rm \, keV}}

\def\GeV{{\rm \, GeV}}

\newcommand{\Hefour}{\ensuremath{{}^4\mathrm{He}}}
\newcommand{\Hethree}{\ensuremath{{}^3\mathrm{He}}}
\nc{\Li}{\ensuremath{{}^7\mathrm{Li}}}

\begin{document}

\title{Cosmological beam dump: constraints on dark scalars mixed with the Higgs boson}

\author{Anthony Fradette}
\affiliation{Department of Physics and Astronomy, University of Victoria, 
Victoria, BC V8P 5C2, Canada}
\affiliation{Perimeter Institute for Theoretical Physics, Waterloo, ON N2J 2W9, 
Canada}
\author{Maxim Pospelov}
\affiliation{Department of Physics and Astronomy, University of Victoria, 
Victoria, BC V8P 5C2, Canada}
\affiliation{Perimeter Institute for Theoretical Physics, Waterloo, ON N2J 2W9, 
  Canada}
\author{Josef Pradler}
\affiliation{Institute of High Energy Physics, Austrian Academy of Sciences, Nikolsdorfergasse 18, 1050 Vienna, Austria}
\author{Adam Ritz}
\affiliation{Department of Physics and Astronomy, University of Victoria, 
Victoria, BC V8P 5C2, Canada}

\date{December, 2018}%

\begin{abstract}
\noindent

Precision cosmology provides a sensitive probe of extremely weakly coupled states due to thermal 
freeze-in production, with subsequent decays impacting physics during well-tested cosmological epochs.
We explore the cosmological implications of the freeze-in production of a new scalar $S$ via the super-renormalizable Higgs portal.
If the mass of $S$ is at or below the electroweak scale, peak freeze-in production occurs during the 
electroweak epoch. We improve the calculation of the freeze-in abundance by including all relevant QCD and electroweak production 
channels. The resulting abundance and subsequent decay of $S$ is constrained by a combination of X-ray data, cosmic microwave background 
anisotropies and spectral distortions, $N_{\rm eff}$, and the consistency of BBN with observations. 
 These probes constrain technically natural couplings for such scalars from $m_S \sim$~keV all the way to $m_S \sim 100\GeV$.  
 The ensuing constraints are similar in spirit to typical beam bump limits, but extend to much smaller couplings, 
 down to mixing angles as small as $\theta_{Sh} \sim 10^{-16}$, and to masses all the way to the electroweak scale. 

\end{abstract}

\maketitle

\section{Introduction}

A pragmatic approach to searching for new physics is to focus on generic interactions that have the potential to be detected experimentally with current or upcoming technology. Classifying the interactions of new neutral states with the Standard Model (SM) according to the dimensionality of the couplings, there are only three `portal' operators that are unsuppressed by a new energy scale. The so-called scalar, vector and neutrino portals could provide the leading connection to a hidden or dark sector, motivated by considerations of neutrino mass and dark matter, but possibly comprising a rich structure of yet-unseen particles and forces~\cite{Essig:2013lka}.

The three portals have recently been under intense experimental scrutiny (see \emph{e.g.} Refs.~\cite{Alekhin:2015byh,Alexander:2016aln}), with a forthcoming program to increase sensitivity into unexplored regions of the parameter space. While collider and beam-dump experiments provide sensitivity to relatively large portal couplings, astrophysical phenomena and cosmology can provide complementary reach to much weaker couplings. Constraints generically arise as follows: thermal production of new states in the very early Universe can occur with a sub-Hubble rate (a process often called `freeze-in'), which necessarily leads to a small but non-negligible abundance of such particles in the thermal bath. If the lifetime of these particles is large, 
they may survive to later epochs, and decay during or after Big Bang Nucleosynthesis (BBN) altering the light element yield. Longer lifetimes 
may lead to decays during or after the formation of the Cosmic Microwave Background (CMB), potentially altering the detailed features observed in
precision CMB experiments.

The origin of these cosmological constraints is reminiscent of the detection strategy behind a generic particle beam dump experiment. Typically, very 
energetic particles in the beam initiate the production of exotic states in the target, which then propagate through a rock or dirt filter, and decay/scatter in a relatively background-free environment inside the detector, thus generating a signal of anomalous energy deposition. In the cosmological setting, the analogue of the initial beam-on-target is the stage of the very early hot Universe, the analogue of propagation through a dirt filter is the long stage of subsequent expansion and cooling, as the Universe evolves into a well-understood stage associated with BBN or the CMB, which is then a direct analogue of the calorimeter-type detector that measures abnormal energy deposition. Therefore, it is
appropriate to name this method of studying rare long lived particles the {\em cosmological beam dump}.

Cosmological constraints of this kind were first applied to the heavy neutral lepton (HNL) portal, and were considered in a number of publications
\cite{Adams:1998nr,Ruchayskiy:2012si,Vincent:2014rja}, resulting in stringent constraints on sterile neutrino degrees of freedom $N$. 
Cosmological constraints on the ultra-weak regime of the dark photon parameter space (a new particle $A'_\mu$ that has an
$\epsilon F'_{\mu\nu} F_{\mu\nu}$ coupling to photons) were explored in Ref.~\cite{Fradette:2014sza} (see also Ref. \cite{Berger:2016vxi}). For the very small coupling constants relevant for cosmological probes, dark photons never thermalize and the abundance is determined by freeze-in production via inverse decay reactions. Subsequent energy injection from dark photon decays can alter the path of BBN and the CMB, and agreement with precision observations excludes certain disconnected regions in the parameter space.
Unlike the case of HNL, thermal production of dark photons may not exhaust all channels, as bosonic states can also be copiously produced during inflation. This extra production channel has, however, a wide range of possible outcomes depending on the Hubble scale during inflation. In that sense, the limits presented in \cite{Fradette:2014sza} are conservative.

A similar study can be carried out for the scalar portal.
Unlike the cases of HNL and dark photons where the leading portal operator is of dimension 4, the neutral scalar $S$ can have a dimension three coupling to the Higgs bilinear, $S H^\dagger H$. 
This represents the only super-renormalizable portal that exists between the SM and any potential dark sector.
Moreover, the radiative corrections to the scalar mass created by such an interaction can be naturally under 
control \cite{Piazza:2010ye}, ensuring technical naturalness of a small $m_S$. 
There has been significant attention paid to this interaction over the last few years, due for example to the idea of 
cosmological self-tuning of the Higgs mass through scanning via the small $S H^\dagger H$ interaction \cite{Graham:2015cka}. 

\begin{figure*}[t]
\centering
 \includegraphics[width= 0.75\textwidth]{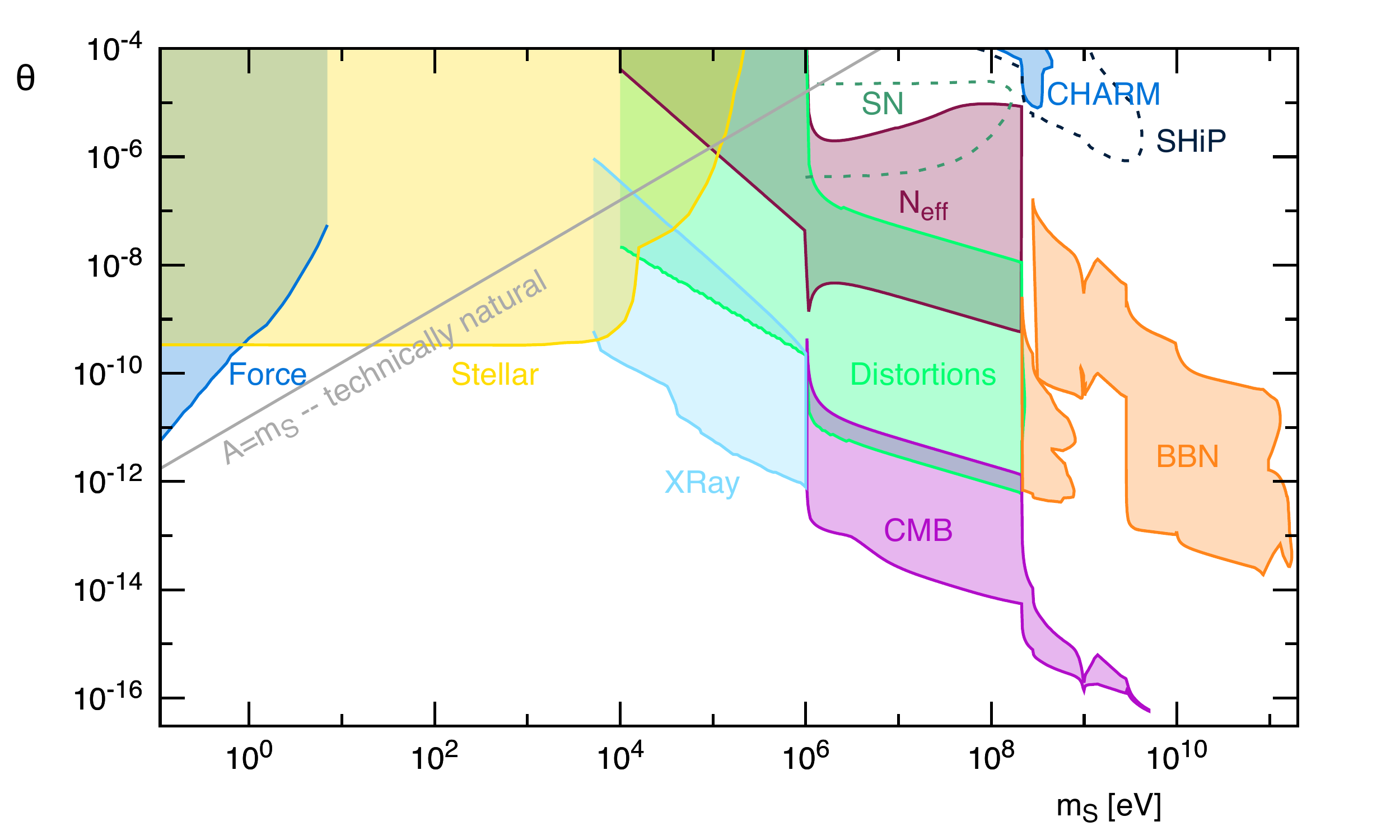}
  \caption{An overview of the excluded parameter space for the super-renormalizable Higgs portal scalar, including the updated constraints from this work due to the diffuse X-ray background (XRay),  CMB anisotropies, spectral distortions, $N_{\rm eff}$ and BBN. Constraints from new short-range forces (Force)~\cite{Kapner:2006si,Decca:2007jq,Geraci:2008hb,Sushkov:2011zz} and stellar cooling (Stellar)~\cite{Hardy:2016kme} from other authors are also shown. We also display the projected SHiP sensitivity~\cite{Alekhin:2015byh} and an estimate of supernova (SN) constraints~\cite{Krnjaic:2015mbs}. %
  }
\label{fig:ParamSpace}
\end{figure*}

With the $S H^\dagger H$ interaction alone, $S$ is guaranteed to be produced in the early universe and its subsequent decay may occur during cosmological epochs that are sensitive to energy injection, as was pointed out in Ref. \cite{Pospelov:2010cw} in an application to BBN. It is therefore interesting in our search for new physics to investigate the phenomenology of this interaction in the super-weak regime, in which the abundance is determined by the freeze-in mechanism. 
The electroweak era can be identified as the main contributor to the freeze-in abundance of $S$, at temperatures where $t(\bar t),\, W^\pm,\,Z,\, h$
are thermally excited, due to the preferential coupling of the SM Higgs to heavy particles.  This was first recognized 
 in \cite{McDonald:2001vt} in the context of the quadratic $S^2 H^\dagger H$ interaction for $m_S > 1\MeV$, but the same conclusion holds for the $S H^\dagger H$ interaction. 
This is markedly different from $N$ and $A'$ freeze-in production \cite{Dodelson:1993je,Redondo:2008ec}, where the peak occurs at noticeably different temperatures $T_{max}$ (for  kinematically accessible particles):
\begin{align}
T_{max}(A') &\propto m_V, \nonumber\\
 T_{max}(N) &\propto O(100\,{\rm MeV}),\\
 T_{max}(S) &\propto M_W. \nonumber
\end{align}

The goal of the present work is to determine the cosmological constraints on $S$, due to its Higgs portal coupling. This requires a computation of the cosmological $S$ abundance due to freeze-in production. Existing estimates of $S$ freeze-in have considered QCD production via top quarks, leading to $Y_S \sim 1.6 \times 10^{12} \theta^2$ \cite{Berger:2016vxi}, where $\theta$ is the mixing angle between $S$ and the SM Higgs, and also a lower bound $Y_S \gtrsim 2.9 \times 10^9 \theta^2$ on the abundance from Primakoff  and Compton processes at low temperatures $T< 20 \GeV$, in the context of relaxion-Higgs mixing \cite{Flacke:2016szy}. (Further in-depth considerations of the freeze-in production due to quadratic and linear couplings of $S$ were performed very recently in 
\cite{Heeba:2018wtf}, and in a more generic setting in \cite{Banerjee:2017wxi}.) Our analysis of the $S$ abundance and decays indicates that QCD and electroweak processes are both significant, and the conclusions are summarized below:
\begin{itemize}

\item \textbf{Freeze-in yield}: The tree-level freeze-in production of $S$ is computed for all electroweak and QCD channels, with a $T$-dependent electroweak vev $v(T)$ used as the first approximation of the relevant thermal effects and to provide an estimate of the precision of the calculation. Solving the Boltzmann equation numerically, and incorporating a full set of QCD cross sections results in a reduction of the total QCD yield relative to the channel analyzed in \cite{Berger:2016vxi}. %
We also assess the accuracy of the Maxwell-Boltzmann approximation in the production calculation, leading to the following result for the abundance from QCD and electroweak processes for the $m_S \ll M_W$ case:
\beq
 Y_S \sim 2.8{-}5.2 \times 10^{11} \theta^2,
\eeq
This estimate becomes more uncertain  for masses $m_S\sim 100\,{\rm GeV}$, i.e. for masses close in value to 
the temperature/energy scale of the electroweak phase transition.

\item \textbf{Decay rate}: We find that there is a sizeable uncertainty in the constraints for $m_S$ in the QCD range, due to the poorly known $S$ decay rate to pions and kaons. We  show two decay profiles in this case. 
Additionally, we improve the calculation of the $S\to \gamma \gamma$ decay rate, which is important for $m_S < 1\MeV$, by incorporating the light quark contribution via mesonic loops, thus decreasing (or increasing) the decay rate by a factor of 4 over the rate used when $u,d,s$ are assumed to be massive (or are neglected).

\item \textbf{Early decays}: Ref.~\cite{Berger:2016vxi} performed a thorough analysis of the BBN constraints, but noted that their analysis of early decays for $m_S < 2 m_\pi$ did not consider energy density considerations due to the large stored energy in the $S$ bath. We include a treatment of early decays, transitioning from the freeze-in abundance to the thermalized freeze-out relic, and consider the impact on the relative neutrino and electromagnetic energy baths.
\end{itemize}

In what follows, we first review the model and describe its features in Sec.~\ref{sec:HiggsPortalModel}. The freeze-in abundance calculation is described in Sec.~\ref{sec:SFreezeIn}, including details of several subtleties. We provide a complete scan of the $S$ parameter space at small mixing angles in Sec.~\ref{sec:CosmoConst}, with details of the cosmological constraints updated in this work. The results are summarized in Fig.~\ref{fig:ParamSpace}, which shows that precision cosmology provides an efficient probe of the
parameter space of the model many orders of magnitude inside the region where it is ``technically natural" ({\em i.e.} not plagued by the issue of fine tuning). Finally, we conclude the paper with a general discussion of the robustness of these results and final conclusions in Sec~\ref{sec:Discuss}. Several technical results are relegated to Appendices.

\section{The super-renormalizable Higgs portal model}

\label{sec:HiggsPortalModel}

We consider a subset of the minimal Higgs portal model, the super-renormalizable Higgs portal. The scalar part of the SM Lagrangian involving the Higgs doublet $H$  is augmented by a mass term for the singlet $S$ and a dimension three interaction:
\begin{align}
\mathcal{L}_{H/S} &\supset \mu^2 H^\dagger H - \lambda_H\left(H^\dagger H \right)^2  - \frac{1}{2} m_S^2 S^2 - A S H^\dagger H. %
\label{eq:L}
\end{align}
The $A$ term induces a small mixing angle $\theta$ between the physical excitations $S$ and $h$. At linear order in $A$, the mixing angle is given by
\beq
\label{mixingT=0}
\theta = \frac{Av}{m_h^2-m_S^2},
\eeq
and leads to Yukawa interactions between $S$ and SM particles, equivalent to the SM Higgs boson interactions rescaled by the suppression factor $\theta$. In the unitary gauge for the broken electroweak phase, after diagonalizing to find the physical states $h$ and $S$, we have the scalar potential
\begin{align}
V_{H/S} &= \frac{m_h^2}{2}h^2 + \frac{m_S^2}{2}S^2 + \lambda v\, h^3 + \frac{\lambda}{4}h^4  \\
&\qquad \qquad + \left(\frac{A}{2}-3\theta \lambda v \right) h^2S  - \theta \lambda \, h^3S, \label{eq:Vhs}
\end{align}
which exhibits the $hhS$ and $hhhS$ contact interactions. 

The $S$ sector could include additional self interactions, \emph{e.g.} $\lambda_3 S^3$. Such self interactions would 
combine with the $A$-term and contribute to the $S$ freeze-in production via Higgs boson decays $h \to SS$. Large self interactions can also influence the $S$ metastable abundance after freeze-in by maintaining thermalization of the dark sector prior to subsequent decays~\cite{Carlson:1992fn,Hochberg:2014dra}. We will neglect this type of interaction and focus on the pure freeze-in regime of Lagrangian~(\ref{eq:L}).

\subsection{$S$ decay rate}

\begin{figure}[t]
\centering
 \includegraphics[width= 0.98\columnwidth]{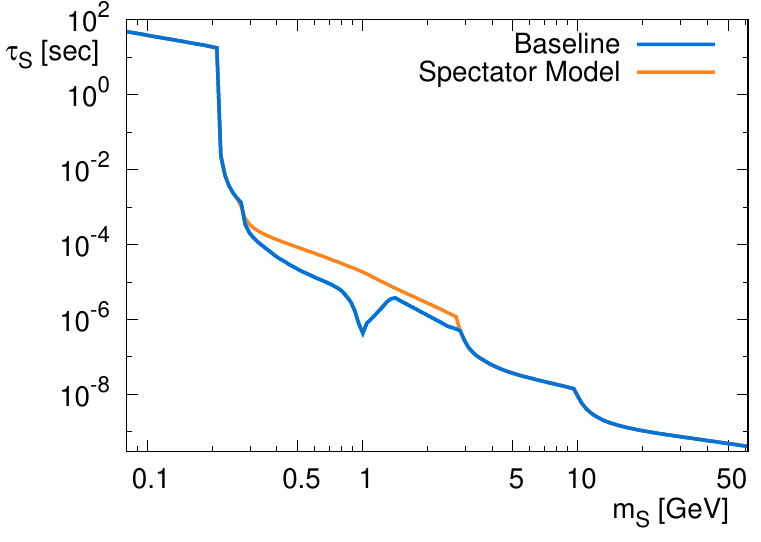}
  \caption{The $S$ lifetime as a function of $m_S$ for $\theta = 10^{-6}$ (reproduced from Ref.~\cite{Fradette:2017sdd}).}
\label{fig:Slifetime}
\end{figure}

The $S$ decay rate has well-known theoretical uncertainties associated with mesonic decay channels in the mass range $2 m_\pi < m_S < 4\GeV$~\cite{Clarke:2013aya}. We follow Ref.~\cite{Fradette:2017sdd} and use two different decay models to demonstrate the magnitude of the theoretical uncertainty in the final $S$ freeze-in parameter space. The baseline decay model matches low-energy theorems near the pion threshold to a $\pi \pi$ phase-shift analysis above 600 MeV by the CERN-Munich group~\cite{Hyams:1973zf} up to $m_S \lesssim 1.4 \GeV$, and interpolates to $m_S \lesssim 2.5\GeV$ where analytical results are expected to be valid~\cite{Bezrukov:2009yw}. For comparison, the spectator model uses perturbative results up to the $c$-quark threshold. In this case, the decay rate into pions is given by low-energy theorems and the kaon or $\eta$ meson contributions are estimated by rescaling the muon branching ratio appropriately~\cite{Gunion:1989we,McKeen:2008gd,Alekhin:2015byh}. The $S$ lifetime in this mass range for the two decay models is shown in Fig.~\ref{fig:Slifetime}. (See also the very recent work \cite{Monin:2018lee}.)

Below the electron threshold, a Higgs-like particle decays to 2 photons through a loop of heavy particles. The leading order decay rate is found by summing over the massive charged particles entering the loop~\cite{Djouadi:2005gi},
\begin{align}
\Gamma (S\to \gamma \gamma) &= \frac{\theta^2 \alpha^2 m_S^3}{256 \pi^3 v^2} \left| C \right|^2, 
\end{align}
where $C$ is a loop function, given explicitly in Appendix A. In this prescription, the light quark degrees of freedom are incorporated through their explicit breaking of chiral symmetry, and the associated mass of pions, kaons and eta mesons~\cite{Pich:1995bw}, i.e. through virtual loops of pions and kaons~\cite{Leutwyler:1989tn}. Adding the contributions from all SM particles, for $m_S \ll 2m_e$ we find
\beq
C = \left\{ 
\begin{array}{llr}
11/3 & \simeq 3.67&\qquad\mbox{for 0 + 6 quarks}  \\
989/522 & \simeq 1.89&\mbox{for 2 + 4 quarks} \\
50/27 & \simeq 1.85&\mbox{for 3 + 3 quarks} \\
1 & = 1&\mbox{for 0 + 3 quarks}
\end{array} \right. 
\eeq
where different scenarios of ($a$ light) + ($b$ heavy) quarks are shown. For the case of 2(3) light flavours 
the pion (pion and kaon) loops are taken into account, while for 0 light flavours they are neglected. 
The true physical value should be close to the $3+3$ or $2+4$ scenarios. Since the difference in decay rate between the two cases, $\mathcal{O}(4\%)$, is negligible for the analysis of new physics, we simply choose $C = 50/27$.

\section{Scalar mixing in the cosmological thermal bath}

In vacuum, the relevant SM masses are generated via the Higgs mechanism and are proportional to the electroweak vacuum expectation value (vev) $v$. In the cosmological thermal bath, and in particular near the electroweak symmetry restoration temperature, long-range interactions are screened by the plasma. Particles effectively develop a thermal mass as a representation of this screening. The mass of a particle at a given temperature $T$ can generally be written as~\cite{Carrington:1991hz}
\beq
m^2(T) = m_0^2(v(T)) + m_T^2(T), \label{eq:mT}
\eeq
where $m_0$ is the zero-temperature mass that depends on the vev and
$m_T$ is the thermal mass. Note that the vev depends on $T$, so that
$m_0$ also has a temperature dependance. A simple analytic formulation
of the high-$T$ Higgs thermal mass parameter in the effective
potential is given by~\cite{Enqvist:2014zqa}
\begin{align}
m_{h,T}^2 (T) &= c_h T^2, \label{eq:mhT}
\end{align}
where
\begin{align}
 c_h&= \frac{1}{16} \left(8\lambda_H + 4 y_t^2 + 3g_2^2 + g_1^2 \right). 
\end{align}
Inserting the additional term $\frac{c_h}{2}T^2 h^2$ into the Lagrangian~(\ref{eq:L}) to generate the equivalent thermal mass, we can solve for $v(T)$
\beq
v(T) = \sqrt{v_0^2 - \frac{c_h T^2}{\lambda_h}}, \label{eq:vT}
\eeq
which predicts an electroweak symmetry restoration at the critical temperature $T_c \simeq 140\GeV$.\footnote{This value is $\mathcal{O}(10\%)$ smaller than the full SM value of $T_c^{\rm SM} \simeq 160 \GeV$ from lattice simulations~\cite{DOnofrio:2014rug,DOnofrio:2015gop}.}
The Higgs thermal mass~(\ref{eq:mhT}) applies to the SM eigenstate prior to mass diagonalization.\footnote{As we are considering \textit{very small} mixing, the effect of $S$ on the SM thermal masses is negligible. Similarly, the $S$ thermal mass will be $m_S^2(T) \sim \theta^2 T^2$ and thus can be neglected for this study. } After this diagonalization of $h-S$ mixing, we obtain a temperature-dependent mixing angle,
\beq 
  \theta (T) = \frac{A\;v(T)}{m_h^2(T) - m_S^2}, \label{eq:thetaT}
\eeq 
which incorporates the leading dependence on temperature for
small $m_S$. However, this expression also signals the presence of a
thermal resonance when $m_S \sim m_h(T)$, which can arise on scanning
$T$ for $m_{h}(T)_{\rm min} \lesssim m_S \lesssim m_{h,0}$. Lattice results indicate
the Higgs thermal mass drops to $m_{h}(T)_{\rm min}\sim15$~GeV at the electroweak crossover \cite{DOnofrio:2015gop}.

\begin{figure}[t]
\centering
 \includegraphics[width= 0.85\columnwidth]{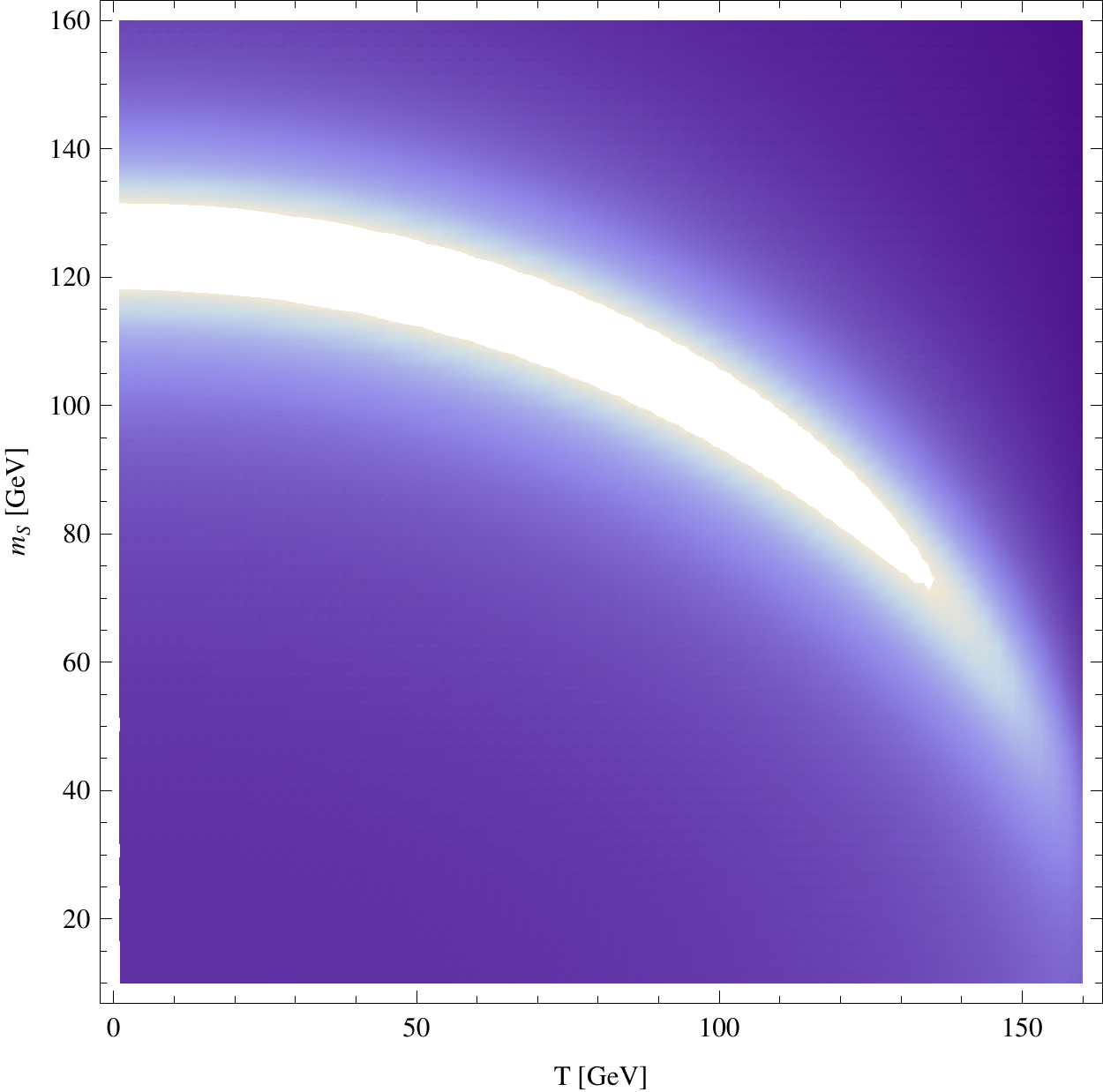}
  \caption{A density plot of the thermal mixing angle $\theta_{\rm eff}(T)$ from (\ref{thetaTeff}), showing the location of the thermal resonance for $T< 140$~GeV. The peak of the resonance defines the resonance temperature $T_{\rm res}$ as a function of $m_S$. The behavior 
  of $m_h(T)$ follows (\ref{eq:mT}) with the naive $v(T)$ model, but with an additional $T$-dependent contribution added to ensures that $m_h(T)$ tracks down to the minimum value  of $m_h(T)_{\rm min} \sim 15$GeV near the crossover transition, as suggested by 
  lattice simulations~\cite{DOnofrio:2015gop}. A finite Higgs damping rate of $0.05T$ was also added for illustration. 
   }
\label{fig:thetaT}
\end{figure}

The apparent divergence in (\ref{eq:thetaT}) at
$m_S = m_h(T)$ is resolved by thermal broadening, which amounts to
replacing the factor of $1/(m_h^2-m_S^2)$ with a Breit-Wigner
propagator for the intermediate metastable Higgs in the rest frame of
the thermal bath. This is conveniently derived by considering the
thermal rate $\Gamma_S$ at which $S$ approaches equilibrium, given by
$\Gamma_S = - {\rm Im}\Pi_S/E$ where $E$ is the particle
energy. $\Gamma_S$ in turn is related to the $S$ production rate $\Gamma_{\rm prod}$ by a Boltzmann factor, 
$\Gamma_S= \Gamma_{\rm prod} - \Gamma_{\rm dest} =
  (e^{E/T}-1)\Gamma_{\rm prod}$. 
  The $S$ self energy takes the form %
\beq 
 \Pi_S(k) = A v(T) \times \frac{1}{k^2-m_0^2(v) -\Pi_h(k)} \times Av(T), 
 \eeq 
 where $\Pi_h$ is the Higgs self energy. Computing
the imaginary part in the on-shell limit, and with
${\rm Re}\Pi_h = m_{h,T}^2$, leads to
\begin{align}
 \Gamma_S &\equiv \theta_{\rm eff}^2(T) \Gamma_h \nonumber\\
     &=A^2\; v(T)^2 \frac{\Gamma_h}{(m_S^2-m^2_{h}(T))^2+(E\Gamma_h)^2},
\end{align}
allowing us to read off the thermally broadened mixing angle, 
\beq
\theta_{\rm eff}^2(T) = \frac{A^2 v(T)^2}{(m_S^2-m_{h}(T)^2)^2+(E\Gamma_h)^2}, \label{thetaTeff} 
\eeq
where $\Gamma_h$ is the Higgs width, or more generically, damping
rate.  The zero temperature width $\Gamma_{h,0} = 4.07 \MeV$ gives a
reasonable approximation for the decay rate, since for the parameter regime of interest here, these decays 
occur late in the cosmological evolution when the temperature is low. Notice, however, that at temperatures around 
the electroweak scale, the damping rate (set by interactions with top quarks and
weak gauge bosons) is expected to scale as
$\Gamma_h(T) \propto T$, and $\Gamma_h(T\sim m_W) \gg \Gamma_{h,0}$.
The effect of thermal broadening at high temperatures can be relevant
for the epoch of freeze-in production.
A density plot of $\theta_{\rm eff}(T)$ from (\ref{thetaTeff}) is
shown in Fig.~\ref{fig:thetaT}. As $T \rightarrow T_c$, simulations
suggest that $m_h(T)$ drops rapidly near
$T_c$ to 15-20 GeV \cite{DOnofrio:2015gop}, potentially allowing a
resonance for any $m_S > 15$~GeV, an effect that is not well captured
by our $v(T)$-scaling model. The potential importance of the thermal
resonance will be considered in more detail in the next section.

We conclude this section by noting an apparent discontinuity in the behaviour of the mixing angle at zero temperature and at temperatures close to the 
phase transition, assuming for simplicity that $m_S$ is a small parameter. From (\ref{mixingT=0}), it follows that $\theta \propto A/(\lambda_H v)$, 
while at finite temperature $\theta_{\rm eff}(T)$, from (\ref{thetaTeff}), scales quite differently as $\theta_{\rm eff}(T) \propto Av(T)/({\rm couplings}\times T^2)$. As $v(T)$ approaches zero, these two formulae have completely different behaviour. 
Taken at face value, this suggests that vertices with Feynman rules
proportional to $v(T)$ will not contribute at all in the electroweak
symmetric phase. However, this is only true at tree-level and the
surviving diagrams are generated at higher order in perturbation
theory and {\em do not vanish} in the limit of $v(T)\to 0$. These
higher order corrections are discussed further in App.~B, and example
diagrams in the higher order expansion are shown in
Fig.~\ref{fig:ZZS_sym}. Therefore, {\em very near} the phase
transition, the thermally corrected mixing angle (\ref{thetaTeff})
will not provide an adequate description of thermal effects, and a
complete treatment of thermal loops would be necessary.

\section{Cosmological production via freeze-in}

\label{sec:SFreezeIn}

The cosmological production rate of a new species $S$, due to $2\to2$ interactions, is given by the Boltzmann equation
\begin{align}
\tilde{s}\dot{Y} &= \int \sum\prod_{i=1}^4 \left(\frac{d^3\nvec{p}_i}{2E_i(2\pi)^3}\right) \Lambda (f_1,f_2,f_3,f_4) |\mathcal{M}|^2 \nonumber \\
&\qquad \times  (2\pi)^4 \delta^4(p_1+p_2-p_3-p_4),
\label{eq:BoltEq2to2}
\end{align}
where $\tilde{s}$ is the entropy density, %
$Y\equiv n_S/\tilde{s}$, while $\Lambda = f_1 f_2 (1\pm f_3)(1\pm f_4)$ represents the thermal distribution of each species and $|\mathcal{M}|^2$ is the spin-summed squared amplitude. In the Maxwell-Boltzmann (MB) approximation for the freeze-in mechanism $\Lambda \to f_1^{\rm MB} f_2^{\rm MB} = e^{-\frac{\left(E_1+E_2\right)}{T}}$. The sum goes over various multiplicity factors, 
such as spin and color.  For 4 different species Eq.~(\ref{eq:BoltEq2to2}) takes the form~\cite{Edsjo:1997bg}
\begin{align}
\tilde{s}\dot{Y}_{12\to3S}&= \frac{g_1 g_2}{8\pi^4}\;T \int_{s_{\rm min}}^\infty ds\;  p_{12}^2 \sqrt{s} \sigma_{12\to3S} K_1\left(\frac{\sqrt{s}}{T}\right)  \label{eq:sYdot},
\end{align}
where $s_{\rm min} = {\rm Max} \left[ \left(m_1 + m_2\right)^2, \left(m_3+ m_4\right)^2\right]$, and 
\begin{align}
p_{12}^2 &= \frac{s}{4} \left(1-\frac{(m_1-m_2)^2}{s}\right)\left(1-\frac{(m_1+m_2)^2}{s}\right), \nonumber\\
\end{align}
while $\sigma = \sigma_{12\to 3S}$ is the standard cross section averaged over initial state degrees of freedom, 
while $g_{1(2)}$ are the spin and color multiplicity factors for initial particles. 

\begin{figure*}[t]
\centering
  \includegraphics[width=.3\linewidth]{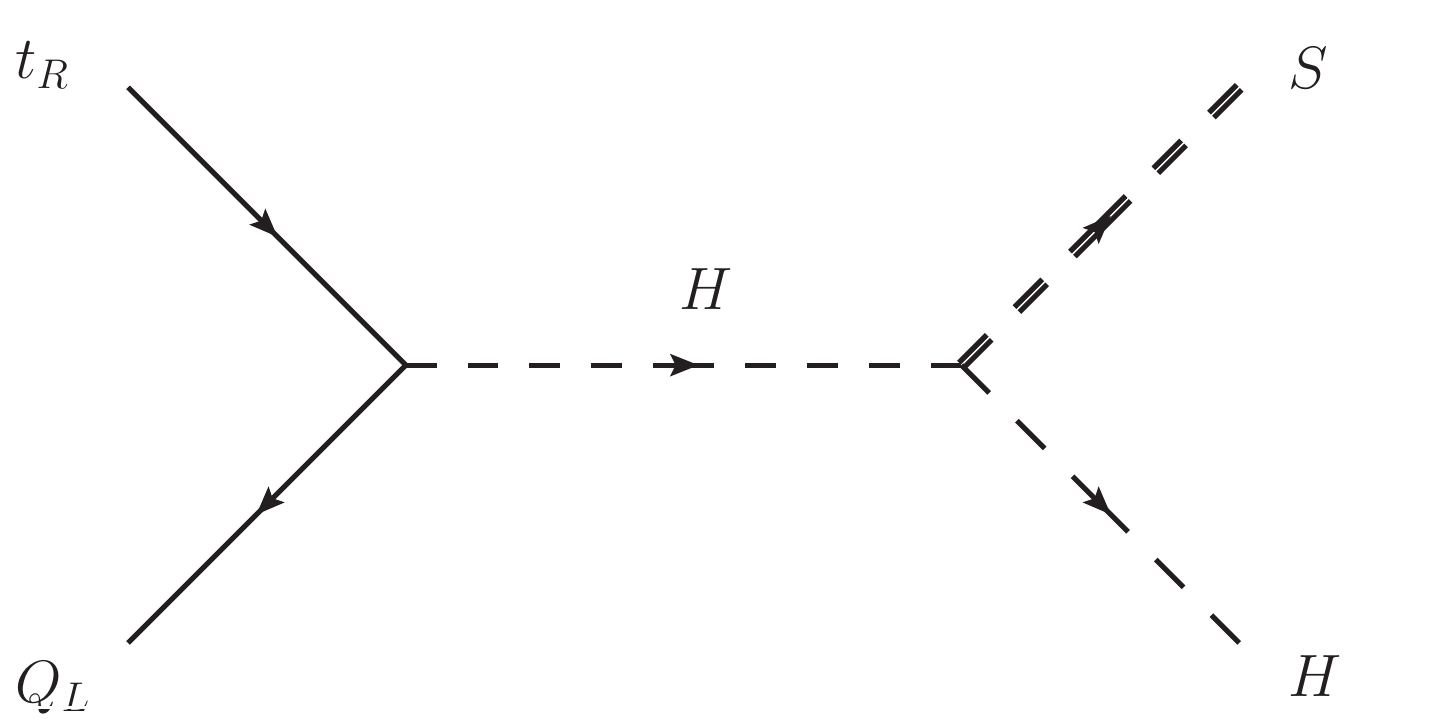}
  \includegraphics[width=.3\linewidth]{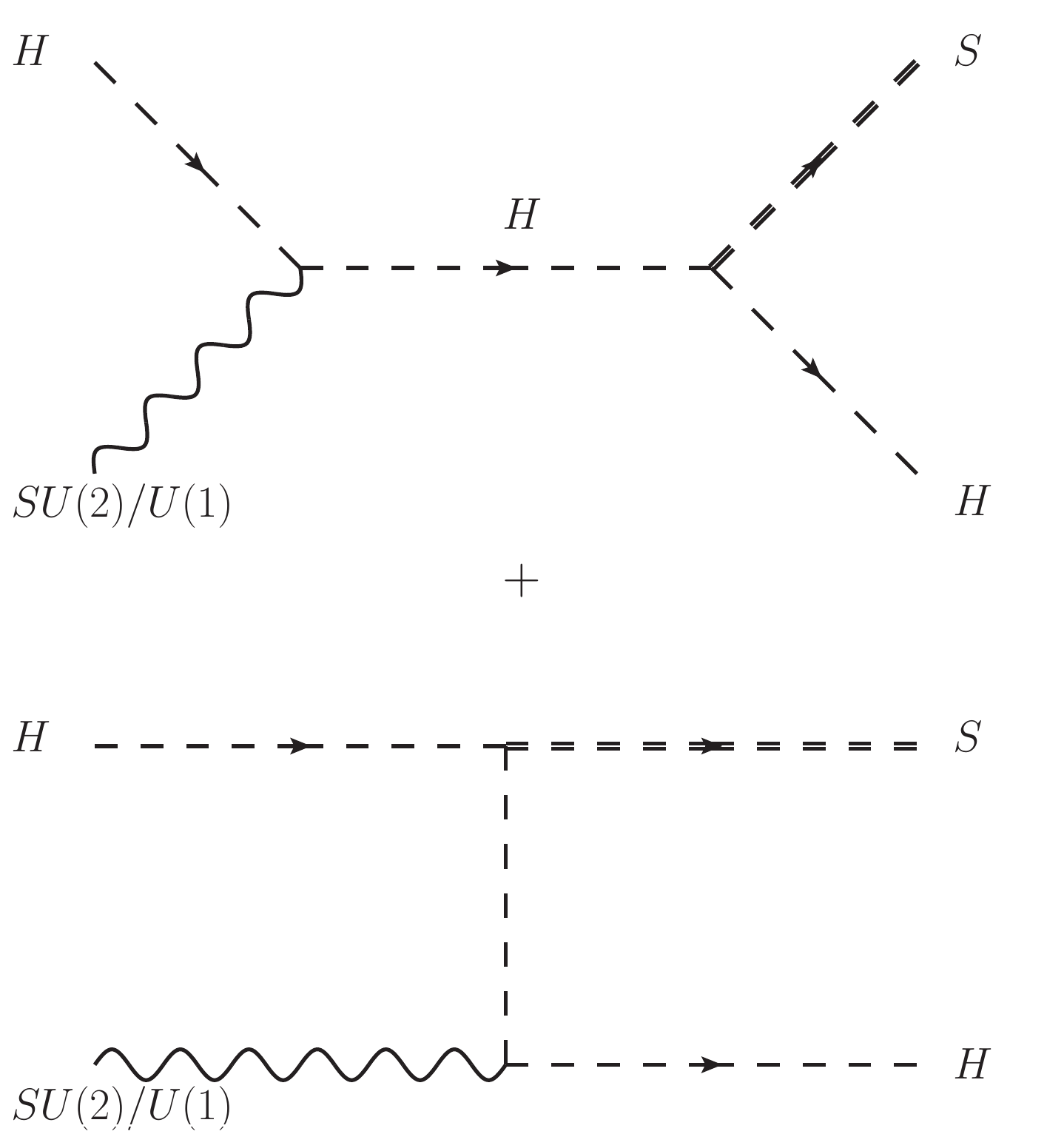}
  \includegraphics[width=.3\linewidth]{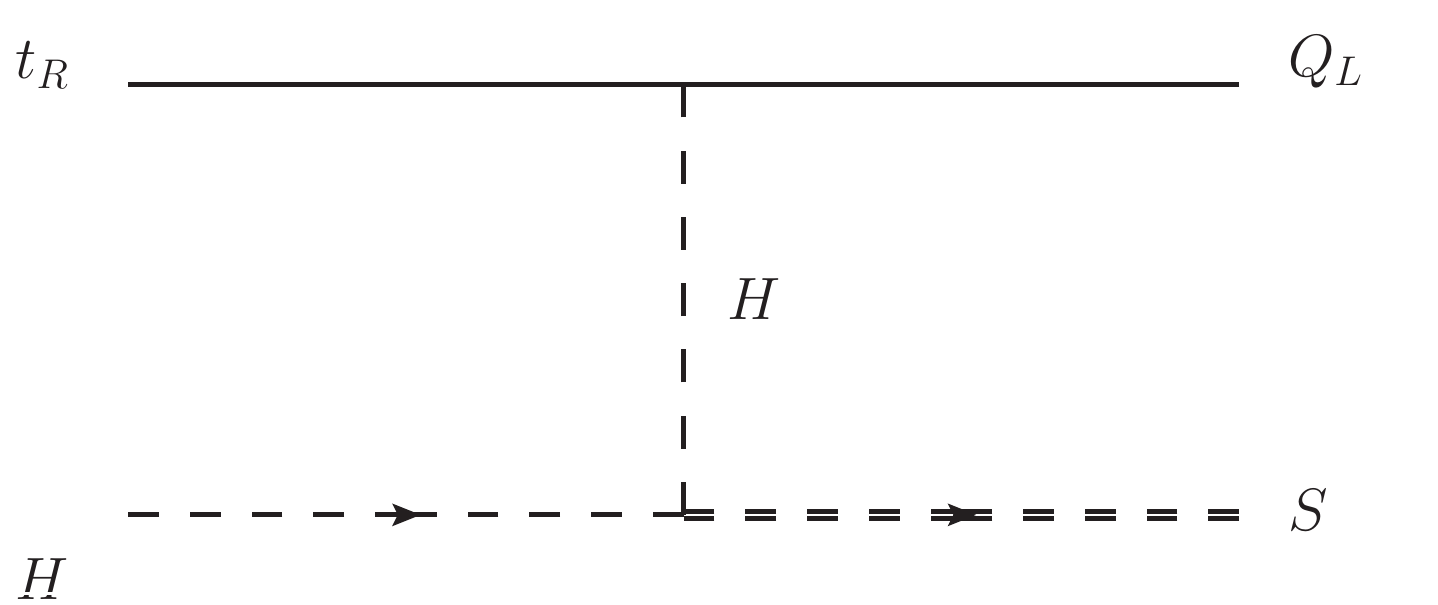}
\caption{$S$-producing interactions in the electroweak symmetric phase. \textit{Left :} Yukawa annihilations. \textit{Center :} Gauge boson scatterings. \textit{Right :} Compton-like scatterings.}
\label{fig:symfeyn}
\end{figure*}

The total $S$ yield is found by summing all possible $12\to 3S$ interactions where 1, 2 and 3 are SM particles. Production channels of the form $12 \to SS$ are suppressed by an extra factor of $\theta$ and are neglected. Since $S$ preferentially interacts with massive particles, we anticipate a large number of possible production channels around the electroweak scale. We classify the different channels by their asymptotic behaviour in the EW unbroken phase. According to the Goldstone boson equivalence theorem~\cite{Cornwall:1974km,Lee:1977eg}, in the $v^2/s\to 0$ limit the behaviour must be determined by the corresponding Goldstone bosons interactions. Expanding the Higgs doublet in the form
\beq
H = \left( \begin{array}{c}
\phi^+ \\ (h + i \phi^0)/\sqrt{2}\end{array} \right),
\label{eq:symH}
\eeq
we find that the only interactions producing $S$ in the symmetric phase will be $2\to 2$ scattering channels, $t_R Q_L \to H S$, $V H \to HS$ ($V$ is a SU(2)/U(1) gauge boson) and $t_R H \to Q_L S$, shown in Fig.~\ref{fig:symfeyn}. We can therefore categorize the $S$-producing interactions as follows:
\begin{itemize}
\item \textit{QCD production}, which includes all diagrams with gluons and top quarks such as $tg\to tS$.
\item \textit{Yukawa annihilation}, which includes the 4 reactions contributing to $t_R Q_L \to H S$.
\item \textit{Compton-like scattering}, which includes reactions with a quark scattering off a boson in the form of $t_R H \to Q_L S$.
\item \textit{Gauge boson scattering}, which includes the reactions purely with electroweak bosons and Higgses.
\end{itemize}

We segment the production calculation into two regimes, first for $T<T_c$ with $v/(v-v(T)) >1$, and then for $T>T_c$, where the vacuum expectation value is negligible and the dimensionful SM couplings proportional to $v$ vanish. $T$ close to $T_c$ 
can be treated by continuity.
In all instances, we compute the cross sections at tree-level, with a few phenomenological improvements justified below. 

In the broken phase, we incorporate the the first thermal corrections by explicitly varying the EW vev as in~(\ref{eq:vT}) and treating all SM masses as vev-dependent variables
\beq
m_{SM}(T) = m_{SM}^0 \times \frac{v(T)}{v_0},
\label{eq:mvT}
\eeq 
and dropping the $T^2$ term in Eq.~(\ref{eq:mT}). As we will see in Sec.~\ref{sec:IRDivs}, we expect this approximation to hold for $v(T)\gtrsim g T$, i.e. until the temperature is high enough that the thermal masses become dominated by plasma contributions.
For $m_h(T)$, the $T^2$ term is retained for consistency in the definition of $v(T)$ and to make sure $\theta (T)$ does not have an unphysical divergence for small $m_S$. %

In the symmetric phase, we retain the quark masses in the cross sections and promote them to thermal masses acquired from the QCD plasma~\cite{Kapusta:2006pm}
\beq
m_q^2 (T) = \frac{g_s^2 C_F}{8} T^2 =  \frac{g_s^2}{6} T^2,
\eeq
which affects the kinematic phase space available for interactions. The Higgs doublet components all obtain the Higgs thermal mass~(\ref{eq:mhT}). We neglect the gauge boson transverse mass. From a finite-temperature point of view, the magnetic thermal mass of a non-abelian SU(N) gauge boson vanishes at one-loop~\cite{Gross:1980br}. A non-vanishing value is generated at higher order as a non-perturbative quantity $m_T^2 \sim (g/3\pi)gT$~\cite{Espinosa:1992kf,Kalashnikov:1991jw}, which is sub-leading compared to the other masses.

In the intermediate regime where a full finite-temperature calculation is needed $\frac{v(T)}{g}\lesssim T \leq T_c$, we extrapolate from the two limiting regimes to obtain an uncertainty band for the model. In either case, we obtain results for 
the relic density that are consistent to within a factor of 2, which is acceptable for the problem at hand.

\begin{figure}[t]
\centering
  \includegraphics[width=.98\columnwidth]{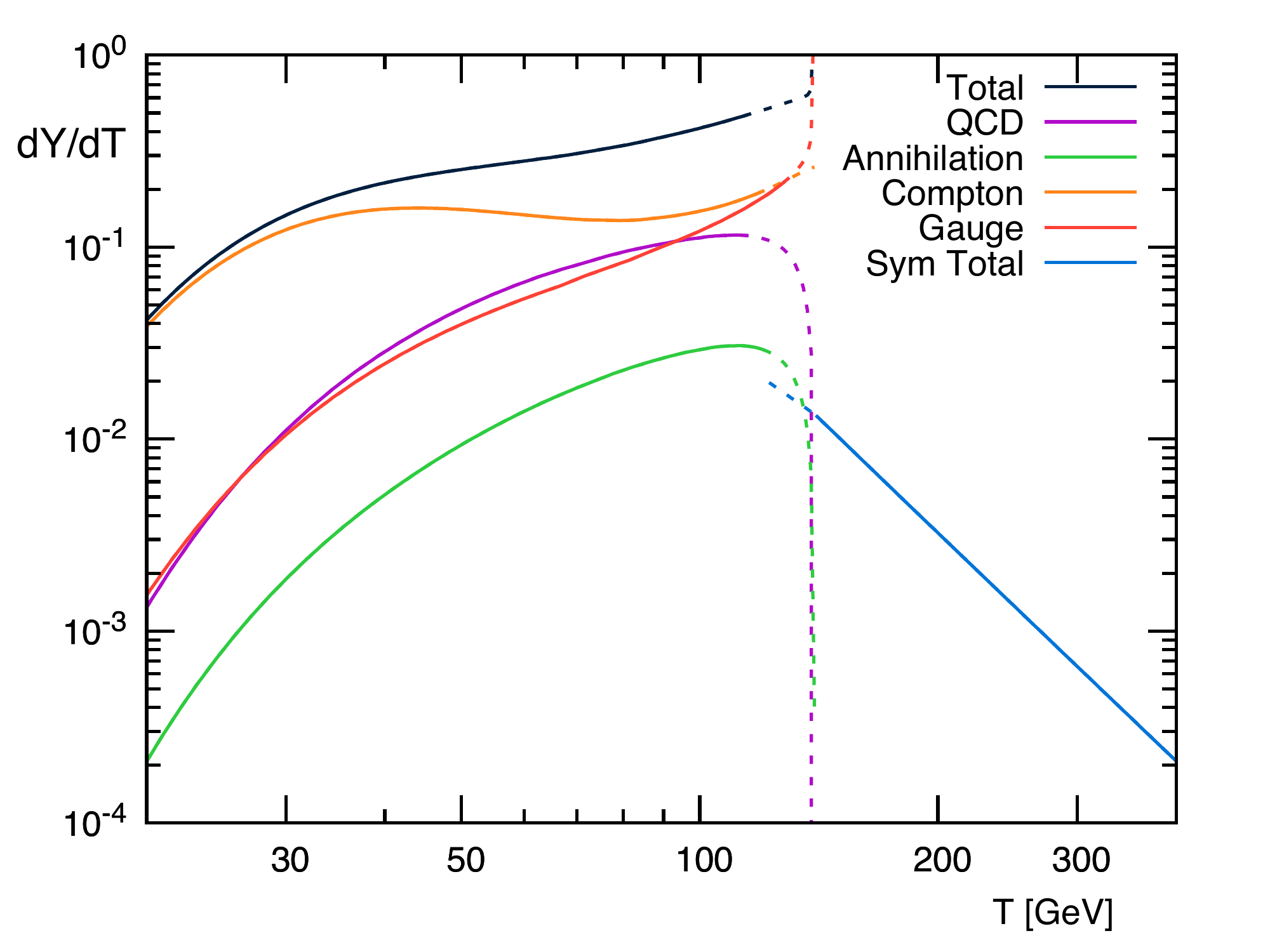}
\caption{Total $S$ freeze-in emissivity and the contribution from each production channel category as a function of temperature for $\theta=10^{-5}$.}
\label{fig:dYdT}
\end{figure}

\begin{figure*}[t]
\centering
  \includegraphics[width=.49\linewidth]{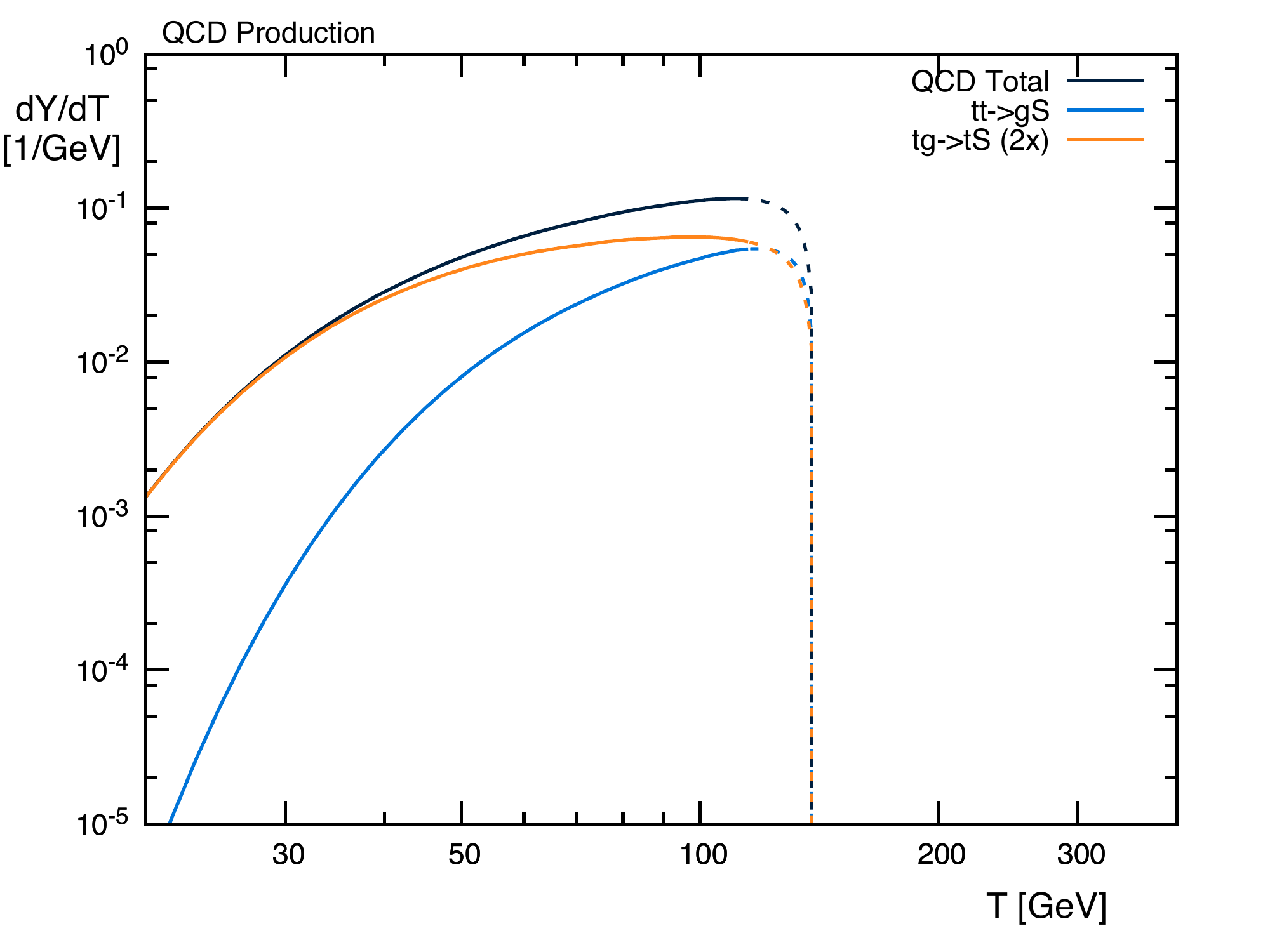}
  \includegraphics[width=.49\linewidth]{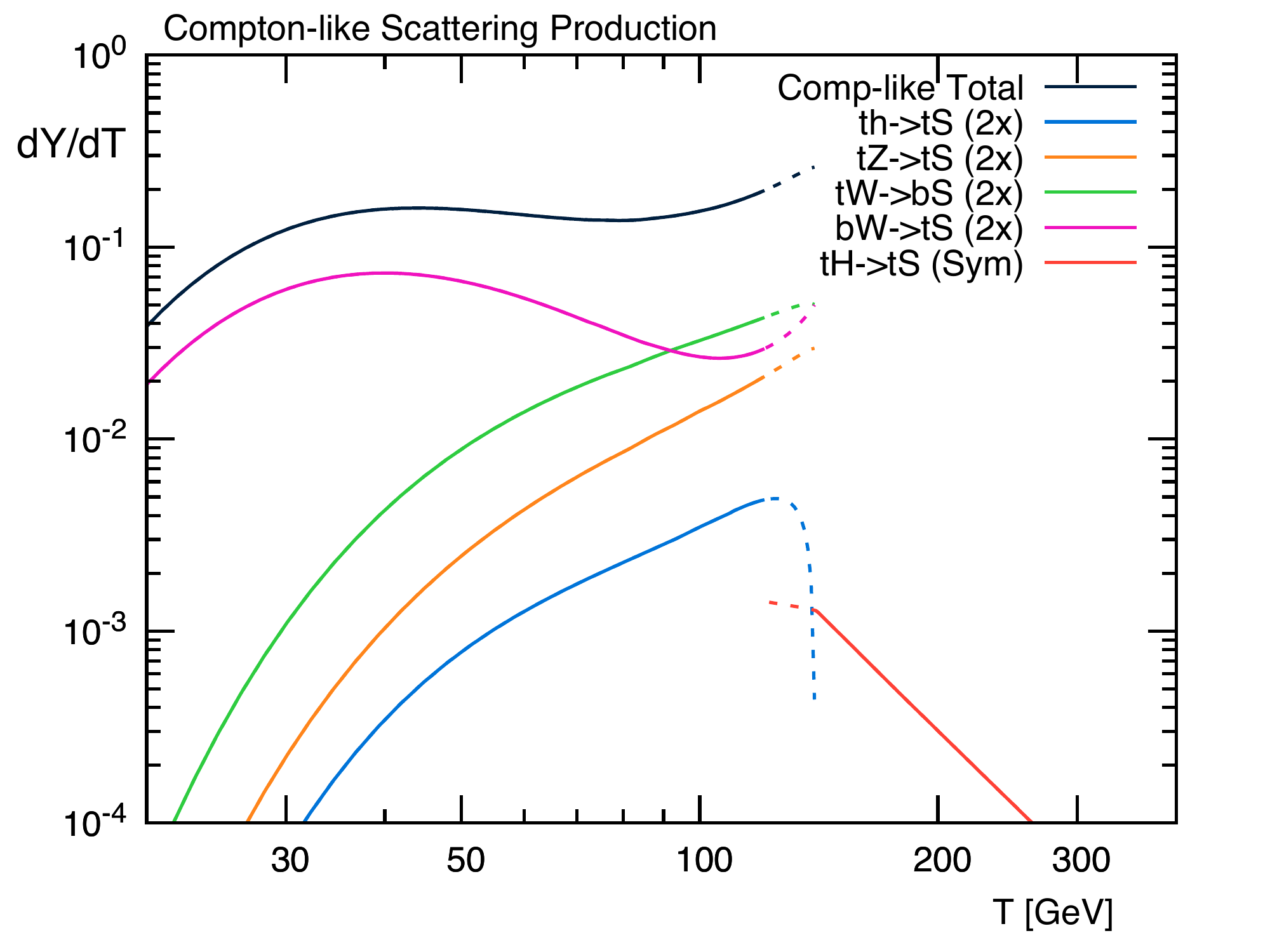}
  \includegraphics[width=.49\linewidth]{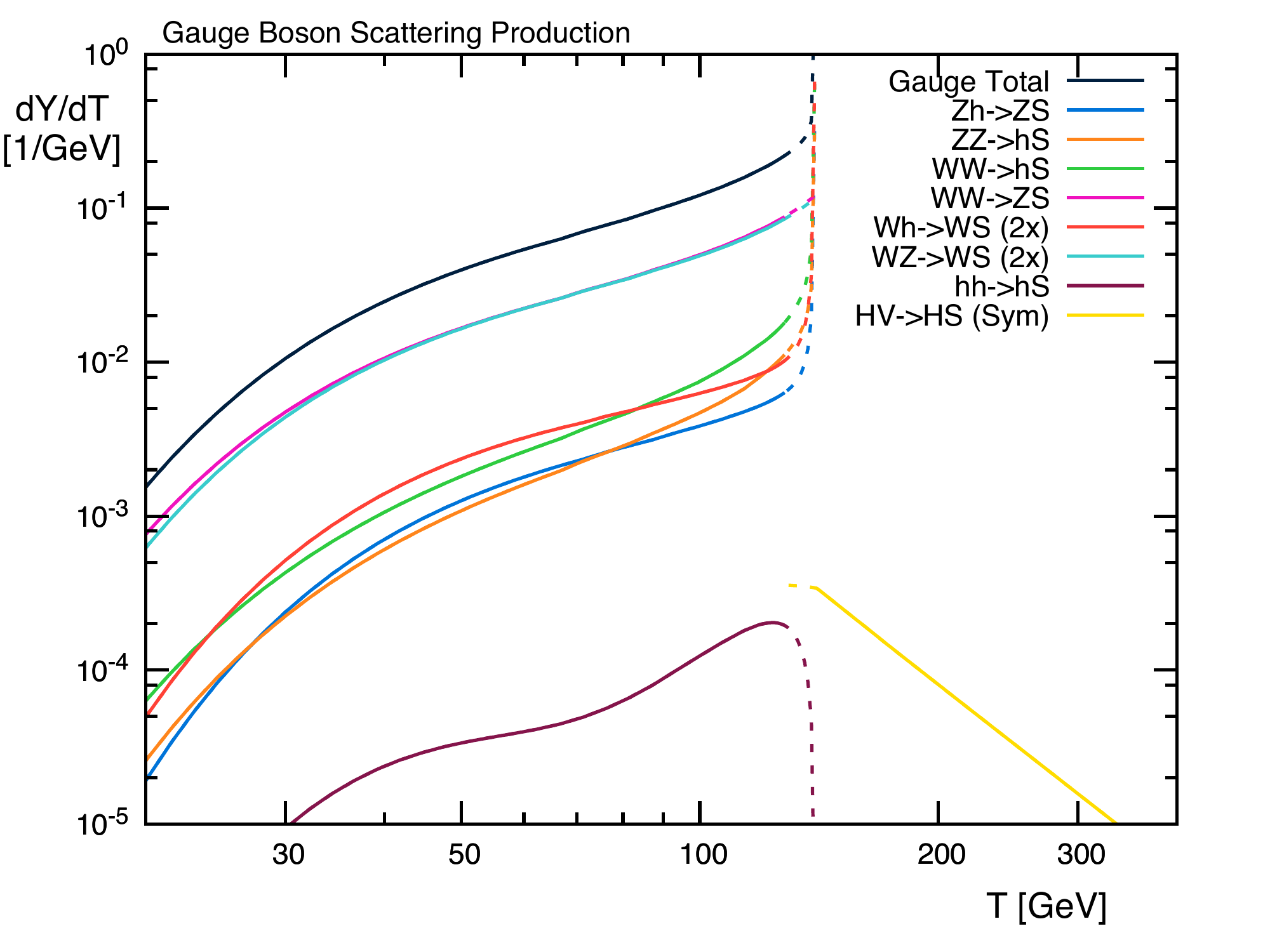}
  \includegraphics[width=.49\linewidth]{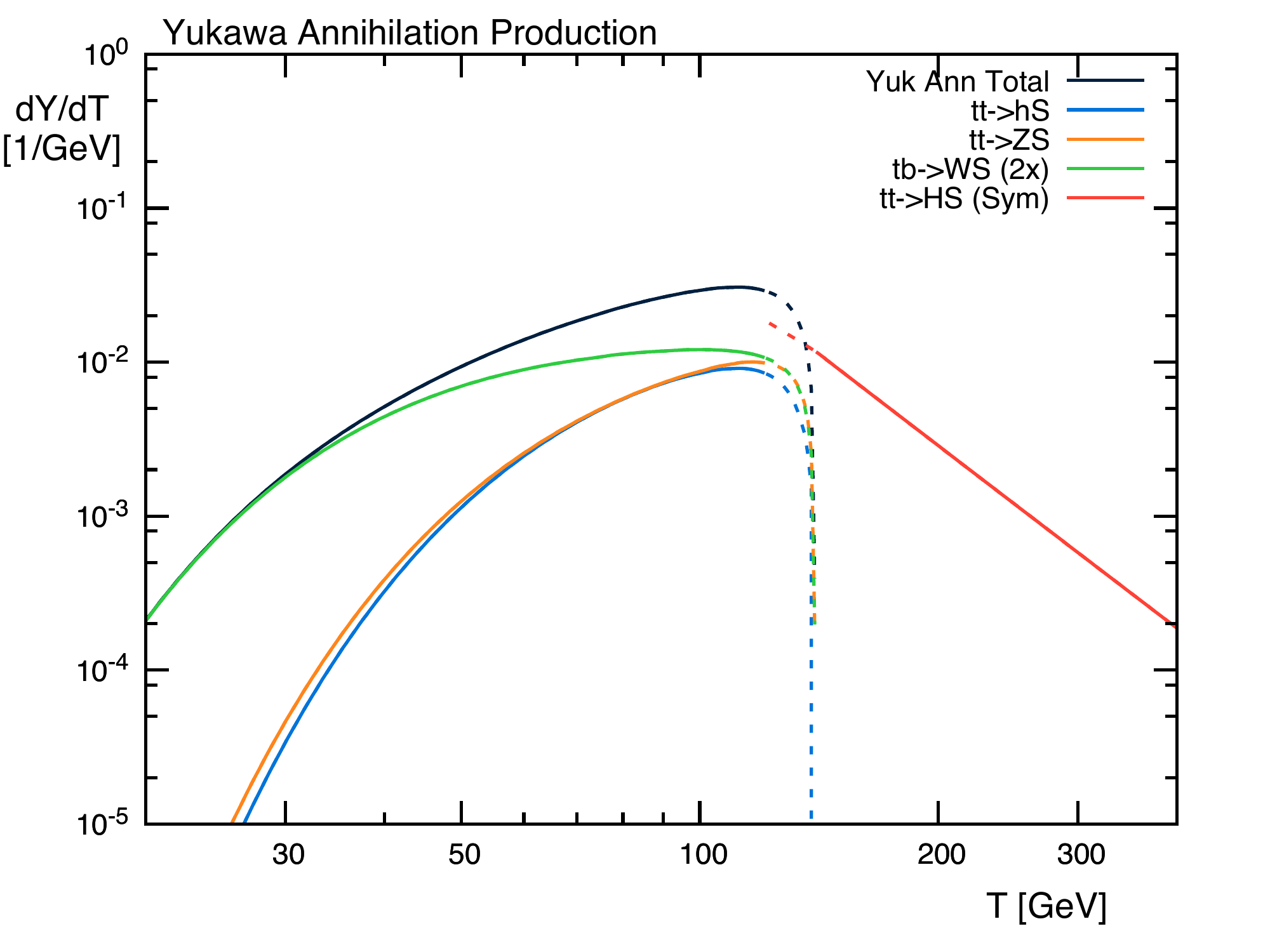}
\caption{The $S$ abundance yield from each production channel category with $\theta = 10^{-5}$. \textit{Top~left~:} QCD production. \textit{Top~right~:} Compton-like scattering. \textit{Bottom~left~:} Gauge boson scattering. \textit{Bottom~right~:} Yukawa annihilation.}
\label{fig:dYdT-indivi}
\end{figure*}

Retaining the top quark Yukawa coupling $y_t$, the electroweak couplings $g,g'$ and the Higgs self-coupling $\lambda_H$ as the only non-zero coupling constants, the yields from each non-vanishing production channel in the $m_S \ll m_h$ limit are compiled in Table~\ref{tab:SYields}. In total, on including the $v(T)$ model, we obtain the following result for the abundance from QCD and electroweak processes:
\beq
 \left.Y_S\right|_{\rm MB\, approx} \sim 3.1{-}3.8 \times 10^{11} \theta^2,
\eeq
for $m_S$ well below the thermal resonance region. The quoted uncertainty band corresponds to whether or not we cut off the production at $T \leq v/g_s \simeq 121 \GeV$ or we push the extrapolation to $T \leq T_c \simeq 140 \GeV$. Later in this section, we will also make an estimate of the precision of the Maxwell-Boltzmann approximation, which will enlarge the precision band somewhat (see Eq.~(\ref{MB})). Note that the total yield from QCD reactions is found to be $Y^{\rm (QCD)}_S \simeq (6.3 - 8.2) \times 10^{10} \; \theta^2$, a factor of approximately $20$ smaller than the value quoted by Berger et al.~\cite{Berger:2016vxi}\footnote{We also point out a disagreement with one of the QCD cross sections quoted by \cite{Berger:2016vxi}, with 
correct expressions given in App. C.}. As listed in Table~\ref{tab:SYields}, we also find many other channels with electroweak gauge bosons that contribute at or above the level of these QCD-induced reactions. We use \textsc{FeynCalc}~\cite{Mertig:1990an,Shtabovenko:2016sxi} to compute the relevant cross sections, and they are listed for completeness in App.~\ref{App:Xsecs}, primarily in the limit of large Mandelstam $s$. The respective emissivity as a function of temperature is shown in Figs.~\ref{fig:dYdT} and~\ref{fig:dYdT-indivi}. 

\begin{table}[b]
\centering
  \begin{tabular}{ | c ||  c | c | c || c |}
    \hline
    Production Channel $i$ &  $Y_i^{v\gg 0}$ &$Y_i^{v\gtrsim 0}$  &$Y_i^{\rm sym}$ & $Y_i^{\rm tot}[10^{10} \theta^2]$\\ \hline \hline
    $t\bar{t} \to gS$  &  2.11 & 0.93 & \multirow{2}{*}{0} & \multirow{2}{*}{6.29{-}8.11} \\ \cline{1-3}
    $t g \to tS$ ($\times$2) & 4.17& 0.90 & & \\ \hline \hline
    $t \bar{t} \to h S$  & 0.41 &0.08& \multirow{3}{*}{0.03{-}0.05} & \multirow{3}{*}{1.72{-}2.01}   \\ \cline{1-3}
    $t \bar{t} \to Z S$  & 0.44 &0.11&& \\ \cline{1-3}
    $t \bar{b} \to W^+ S$ ($\times2$) & 0.82 &0.11&&  \\ \hline\hline
    $th \to t S$ $(\times 2)$ & 0.38 & 0.13  & \multirow{4}{*}{0.14{-}0.21} & \multirow{4}{*}{14.40{-}17.77}  \\ \cline{1-3}
    $tZ \to t S$ $(\times 2)$ & 1.46 & 0.77& &   \\ \cline{1-3}
    $tW \to b S$ $(\times 2)$ & 3.66 & 1.43& &   \\ \cline{1-3}
    $bW \to t S$ $(\times 2)$ & 8.70 & 1.11 && \\ \hline \hline
     $Zh\to ZS$  & 0.26&0.10 &\multirow{7}{*}{0.01{-}0.02} & \multirow{7}{*}{8.68{-}10.93}  \\ \cline{1-3} 
          $ZZ\to hS$  & 0.33 & 0.17 &&\\ \cline{1-3} 
          $WW\to hS$ & 0.57 & 0.25  && \\ \cline{1-3}  
     $WW\to ZS$ & 3.47 & 0.89 &&  \\ \cline{1-3}  
     $Wh\to WS$  $(\times 2)$ & 0.46 & 0.16 &&  \\ \cline{1-3}  
     $WZ\to WS$  $(\times 2)$& 3.57 & 0.69 &&  \\ \cline{1-4}  
   $hh \to h S$ & 0.01  &$ <0.01$ &0&  \\ \hline \hline \hline
    Total & 30.81 &7.84  &0.19{-}0.28&31.1{-}38.8  \\ \hline 
   \end{tabular}
\caption{$S$ freeze-in yields for small $m_S$. For each production channel, the yield is given in units of $10^{10}\theta^2$, separated into the near-vacuum contribution $Y_i^{v\gg 0}$ and the additional yield $Y_i^{v \gtrsim 0}$ if extrapolated to the phase transition. The yield from each production category in the symmetric phase is shown as $Y_{\rm cat}^{\rm sym}$ with the range displaying the total yield for $T$ above $T_c$ or extrapolated down to $T \gtrsim g v(T)$.}
   \label{tab:SYields}
\end{table}

\begin{figure}[t]
\centering
  \includegraphics[width=.98\linewidth]{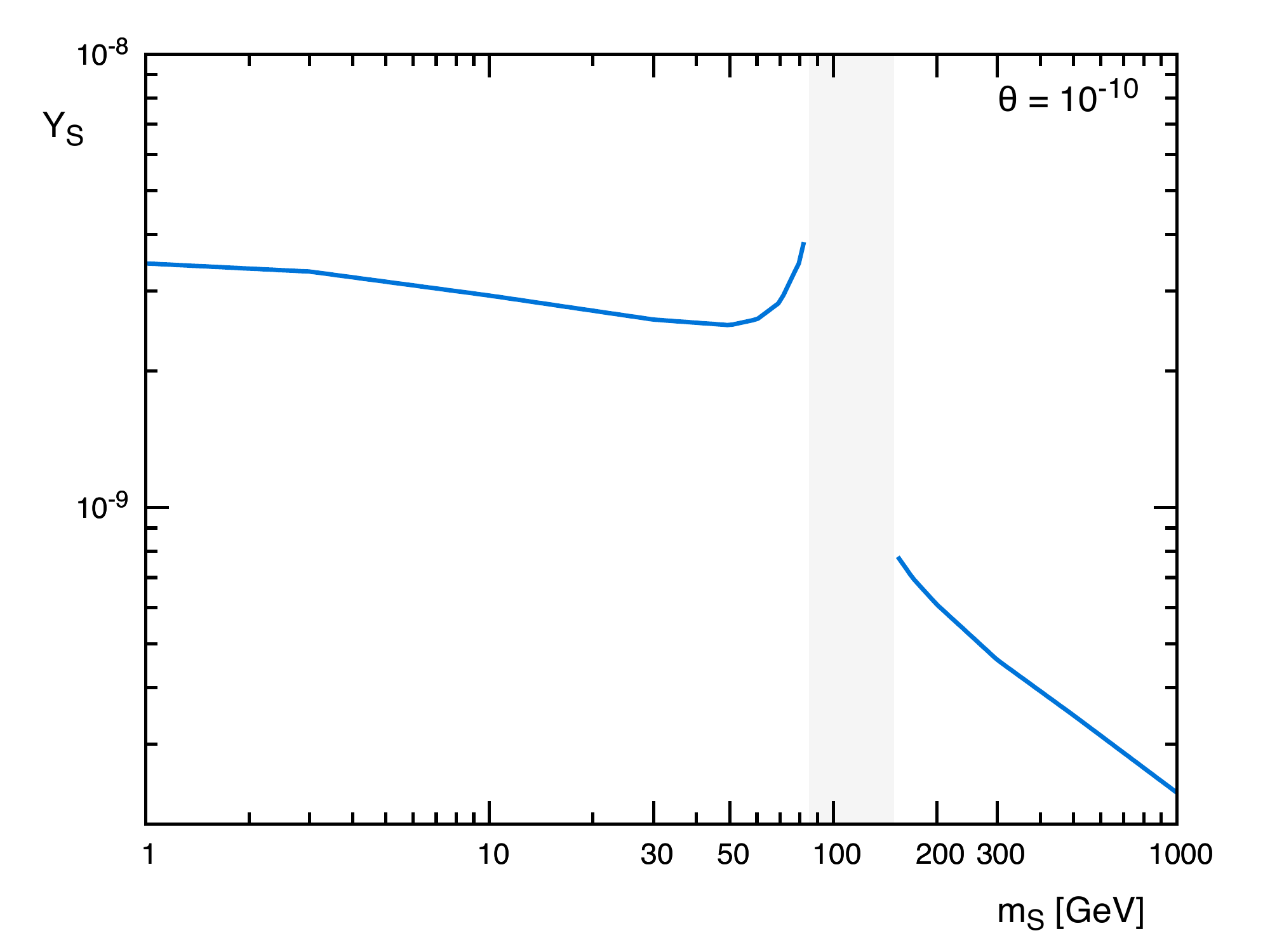}
\caption{The total $S$ abundance yield from non-resonant $2\rightarrow 2$ production channels as a function of $m_S$ (with $\theta = 10^{-10}$). The mass region where $S$ is resonant with the Higgs boson is excluded, and we note that for $m_S$ above a few tens of GeV, other production channels may also be significant. (See the text for details.)}
\label{fig:YS}
\end{figure}

Thus far, we have discussed the parameter range where $m_S$ is parametrically smaller than the weak scale. In this regime, 
with production occurring mostly at weak scale temperatures, the prediction for $Y_S$ is approximately independent of $m_S$. However, this will change once $m_S$ is
 increased to a scale 
comparable to the thermal masses of particles in the SM bath. In Fig.~\ref{fig:YS}, we plot the 
resulting value of $Y_S(m_S)$ that follows from generalizing the $2\to 2$ production 
mechanisms discussed above to finite $m_S$. We note that once 
$m_S$ reaches a few tens of GeV, the $2\to 2$ production channels
may no longer be dominant, and other processes such as resonant oscillations ($1\to1$ production), and inverse 
decays ($2\to1$ production), may also contribute significantly to $Y_S$. We discuss such contributions in separate subsections below.

\subsection{Infrared divergences}

\label{sec:IRDivs}

\begin{figure}[t]
\centering
  \includegraphics[width=.98\columnwidth]{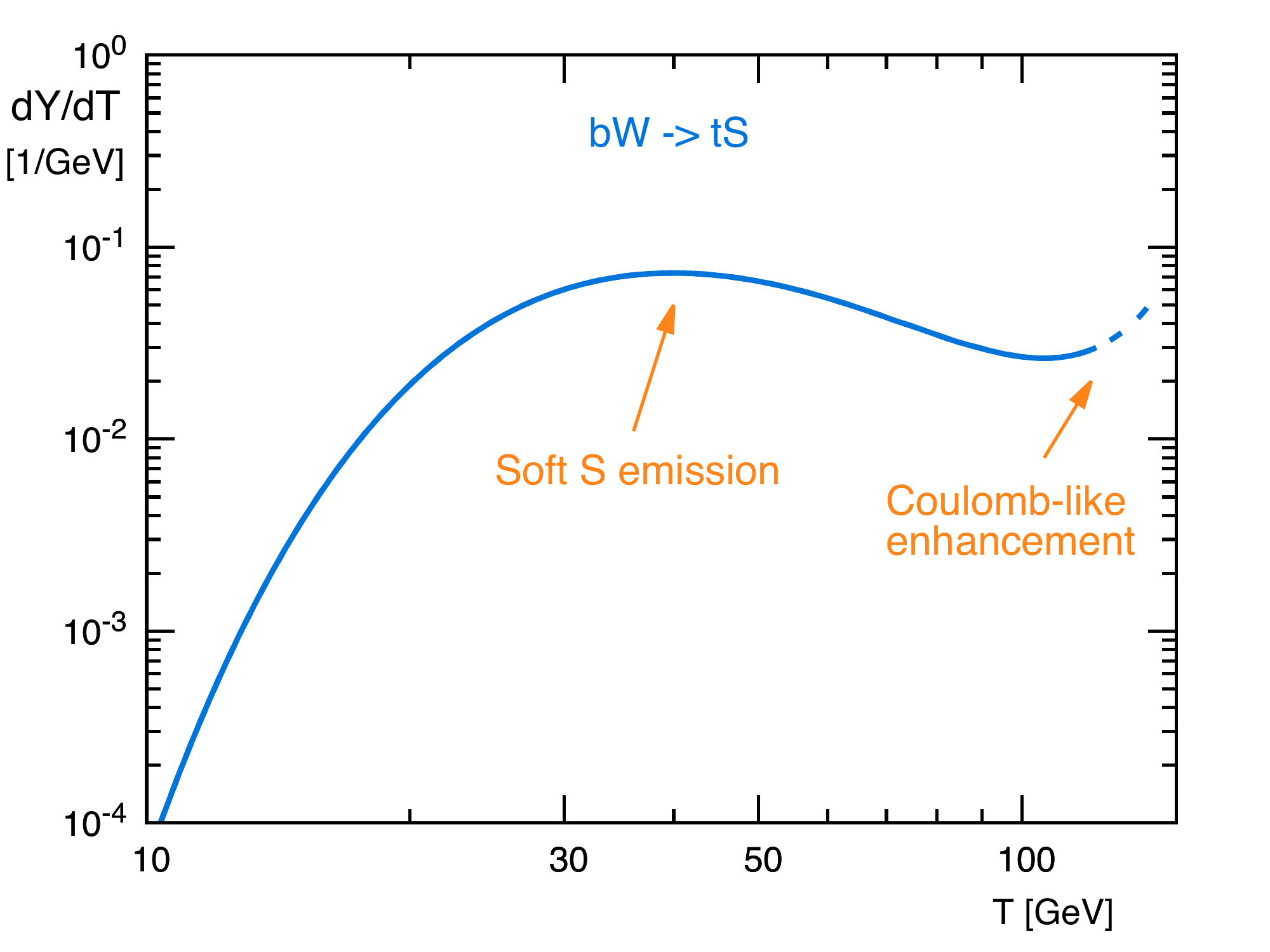}
\caption{ The emissivity of the production channel $bW\to tS$ showing the two types of IR divergences present in the calculation. The soft $S$ emission is physical and unique to this production channel. The Coulomb-like enhancement is present in all reactions with a $t$ or $u$-channel spin-1 mediator, is unphysical and signifies the breakdown of our calculations. (See the text for details.)}
\label{fig:IRdivs}
\end{figure}

When calculating various $2\to2$ production processes using the simplified $v(T)$ 
approach, we encounter additional complications due to the infrared sensitivity of the 
production cross sections. There are two types of infrared divergences in the interactions that require special care, both present in the channel with the largest yield 
contribution, $b W \to t S$, schematically shown in Fig.~\ref{fig:IRdivs}. 

At lower temperatures, where the vacuum cross sections are clearly applicable, the emission of a soft $S$ is enhanced by the near-on-shell $t$ and $W$ mediators. Removing the $S$ emission, the inverse decay process $bW \to t$ is kinematically allowed. Since we are considering an arbitrarily light scalar, $m_S \ll m_W, m_t$, the $2 \to 2$ reactions creating the $S$ have a kinematic cutoff $s\geq (m_t + m_S)^2$, which approaches the propagator singularity $1/(s-m_t^2)$ as $m_S \to 0$. This type of divergence is regulated by the finite width of the propagator. For this channel, we promote the denominator of the $t$ quark and $W$ boson propagators to their Breit-Wigner equivalent
\begin{align}
\frac{1}{p^2-m_{t/W}^2} &\to \frac{1}{p^2-m_{t/W}(T)^2 - i \Gamma_{t/W}(T) E_{t/W}}, %
\label{eq:Breit-Wigner}
\end{align}
where $\Gamma_i(T) = \Gamma_i^0 \times \frac{v(T)}{v_0}$ is consistent with our $v(T)$ model throughout the calculation and the SM values are $\Gamma_t^0 = 1.4 \GeV $ and $\Gamma_W^0 = 2.1 \GeV$~\cite{Patrignani:2016xqp}. The resonances are 
further broadened by thermal effects. Multiple schemes for calculating cross sections with unstable particles have been proposed beyond the simple substitution~(\ref{eq:Breit-Wigner}). The basic Breit-Wigner is technically incompatible with gauge invariance and Ward identities~\cite{Argyres:1995ym,Kauer:2001sp,Schwinn:2003fp}, a problem that can lead to dramatic inconsistencies in the small-angle scattering away from the resonance~\cite{Argyres:1995ym}. Explicitly comparing the 
cross section with and without the substitution~(\ref{eq:Breit-Wigner}) away from the resonance, we find
\beq
\frac{\sigma_{bW\to tS}^{BW}}{\sigma_{bW\to tS}^{0}} \xrightarrow{s\gg m_t^2,m_W^2} \frac{m_W}{\Gamma_W}\arctan{\frac{\Gamma_W}{m_W}} =0.9998
\eeq
which justifies the use of Breit-Wigner propagators in this reaction near the singular point. 

The second class of infrared divergences appear near $T_c$, where $v\to 0$. The exchange of a massless spin-1 particle in the $t(u)$-channel generates a well-known collinear divergence in forward (backward) scattering~\cite{Berestetsky:1982aq}. Therefore, 
with the approximation $m_W^2 = \frac{1}{4}g^2v^2(T)$, when the mass of $m_W$ approaches zero, all the
$t$-channel exchange diagrams are necessarily enhanced. Once again, thermal effects will come to the rescue and 
stabilize these divergences.

In the context of $S$ freeze-in production, the total cross sections with the $t$ or $u$-channel gauge boson propagators do not fall off as $1/s$ in the high energy limit (see App.~\ref{App:Xsecs} for expressions). For the example shown in Fig.~\ref{fig:IRdivs}, we have
\beq
\sigma_{bW^+\to t S} \to \frac{\theta^2 m_W^2}{12 \pi v^4} + \mathcal{O}\left(1/s\right)\nonumber\\
\to \frac{\theta^2 g^2}{48 \pi v^2} + \mathcal{O}\left(1/s\right).
\label{eq:bWtS}
\eeq
At low temperatures, this asymptotic behaviour does not matter as large values of $s$ are exponentially suppressed by the energy available in the initial particle distributions, %
which clearly diverges as $v(T)\to 0$. Conceptually, this IR divergence should be regularized in the same fashion as the scalar QED example. In particular, $g^2v^2(T)/4$ will receive an additional $m_{W,T}^2(T)\propto g^2T^2$  temperature-dependent correction. 
Thus the Coulomb-like enhancement near $T\to T_c$ obtained from a simple extrapolation of the vacuum cross sections, with vev-dependent masses (as in Eq.~\ref{eq:mvT}), signifies the breakdown of our calculation as thermal effects are not incorporated. The formal strategy to deal with collinear IR divergences in thermal field theory has been laid out by Braaten and Yuan~\cite{Braaten:1991dd} in the weak coupling limit $g\ll 1$. 
We will not perform this full calculation, but simply use the limit $gT \lesssim v(T)$ as the boundary of validity for our tree-level cross sections.

\subsection{Resonant $S$ production}

\label{sec:Sres}

As seen in Fig.~\ref{fig:thetaT}, the mixing angle has a physical resonance when $m_h(T) \simeq m_S$. Near resonance, 
the $h\to S$ oscillation may become efficient, and contribute to the overall yield $Y_S(m_S)$. 
Below, we are going to show that the contribution of the resonance is not important for production of very light $S$ particles, 
while it can contribute significantly, starting at $m_S$ in the range of few tens of GeV. 
In practice, there is a significant uncertainty in the behaviour of $m_h(T)$, and the lowest value $m_h(T)$ can acquire as  
a function of temperature.
Recent lattice simulations suggest that near $T_c$ the thermal Higgs mass $m_h(T)$ drops rapidly to 15-20 GeV, in a manner reminiscent of a second-order phase transition. In principle, this allow the resonance to arise for any $m_S$ above this scale. Physically, the resonance arises when the virtual Higgs that rotates into $S$ is allowed to go on-shell, and the corresponding mixing angle develops a Breit-Wigner form (\ref{thetaTeff}) associated with the Higgs thermal width.

Although our primary interest is in lower values of $m_S$, where the resonance is not present, it is interesting to consider the enhancement associated with the thermal resonance. Since $\Gamma_h \ll m_h $ even after accounting for thermal broadening, we can use the narrow-width approximation (NWA) to estimate the $S$ yield from the resonance. Taking the NWA in (\ref{thetaTeff}), we obtain 
\beq
\theta^2_{\rm eff} \to \theta^2_{\rm NWA} = \frac{A^2v^2(T_{\rm res})\pi}{2m^2_S E\Gamma_h(T_{\rm res})|m_h'(T_{\rm res})|} \delta \left(T-T_{\rm res}\right).
\eeq
where $m_h'(T_{\rm res})$ is the temperature derivative of $m_h(T)$ evaluated at resonance.
Substituting into (\ref{eq:sYdot}), we find the simplified integral
\begin{align}
Y_{S,{\rm res}} &\simeq \int \frac{dn_h(E,T_{\rm res})}{s(T_{\rm res})H(T_{\rm res})T_{\rm res}} \frac{\pi A^2 v^2(T_{\rm res})}{m_S E |m'_h(T_{\rm res})|},
\label{Yres}
\end{align}
where $dn_h(T,E)$ is the Bose-Einstein distribution for the Higgs boson. %
Notice that the overall damping rate $\Gamma_h$ for the Higgs boson drops out of this formula, 
and its main uncertainty is encapsulated in the value for $T_{\rm res}$ and $m'_h(T_{\rm res})$. 
Evaluating the remaining integral (and using simplified Maxwell-Boltzmann statistics in the process) we arrive at an analytic  estimate for the $h\to S$ oscillation-induced abundance,
\begin{equation}
\label{resresult}
Y_{S,\rm res} \simeq \left.\frac{\pi\theta_{\rm vac}^2}{2(2\pi)^3}\frac{(m_{h,\rm vac}^2-m_S^2)^2v(T)^2K_1(m_S/T)}{v^2H(T)s(T)|m'_h(T)|}  \right.,
\end{equation}
where all thermal quantities are to be evaluated at $T= T_{\rm res},\,
m_h(T_{\rm res}) = m_S$. 

The simplicity of (\ref{resresult}) is deceiving. Depending on the assumed behaviour of $m_h(T)$, the results can vary substantially. It is possible, however, to conclude that 
if one takes the most extreme behaviour, $m_h(T) = (v(T)/v) \times m_{h,\rm vac}$, which should conservatively over-estimate resonant contribution, the result is still quite small 
for small $m_S$. In particular, we find
\begin{equation}
Y_{S,\rm res}(m_S < 2 m_b) \leq 10^{10} \theta^2.
\end{equation}
On the other hand, for $m_h \simeq 100$ GeV, our results indicate that $Y_{S,\rm res}$ can reach $\sim 4 \times 10^{11}\theta^2$
and become comparable, or even larger than the non-resonant contributions. Interestingly, for $m_S$ as high as 100 GeV, the 
uncertainty in the resonant contribution becomes {\em smaller}, due to the fact that the resonance occurs at temperatures significantly lower than the cross over temperature.) 

Another interesting observation is that for $m_S \simeq m_h$ the actual cosmological constraints are weaker than for 
$m_S \neq m_h$. In that limit, Eq. (\ref{resresult}) is not applicable, and one has to retain the proper thermal damping rate for the production calculation. The point is that all constraints are very asensitive to lifetime of $S$, and the effect of close-to-resonance 
mixing on the decay rate is very pronounced, leading to significant shortening of
 lifetime, and relaxation of the bounds despite enhanced production. 

To ensure that our constraints are conservative, we focus on $m_S$ below the weak scale and do not include the resonant contribution to $Y_S$.

\subsection{Production via Inverse Decays at large $m_S$}

We have concentrated on $2\to 2$ and resonant production modes of the
$S$ scalar, which are dominant for $m_S$ below the weak
scale. However, at much larger values of $m_S$, there is also an inverse decay, or
$2\to 1$ type production channel, that we comment on briefly in this
subsection.

The treatment is simplest when the mass of $S$ is asymptotically
larger than the weak scale.  The decay is then predominantly to
longitudinal $WW$, $ZZ$ and to $hh$ pairs, or equivalently into four pairs of
real scalars.  The total width is
\begin{equation}
\Gamma_S = \frac{A^2}{8\pi m_S}= \left.  \frac{\theta^2m_S^3}{8\pi v^2}\right|_{m_S \gg m_h}.
\end{equation}
Production is governed by the very same width, and appropriately
modifying previous results for dark photons \cite{Fradette:2014sza},
we obtain the corresponding estimate for the yield,
\begin{equation}
Y_S = \frac{3}{4\pi} \times \frac{\Gamma_S m_S^3}{(Hs)_{T= m_S}}.
\end{equation}
Parametrically, this result scales as $\theta^2{M_{\rm Pl}  m_S}/v^{2} $ or ${M_{\rm Pl} A^2 }/m_S^{3}$, 
where $M_{\rm Pl}$ is the Planck mass, while numerically we find,
\begin{equation}
Y_{S,{\rm ID}}(m_S={1\,\rm TeV}) \simeq 2.5 \times 10^{12} \theta^2.
\end{equation}
Notice the slightly larger overall numerical coefficient, which results from less phase space suppression for the inverse decay 
process compared to $2\to2$ processes.

Jumping forward to consider potential cosmological sensitivity in this high $m_S$ regime, we note that the best chance of constraining the model is provided by
BBN (as the lifetime is too short for other probes). 
Normalizing the decay width to the most sensitive lifetime window,  $\Gamma_S = 1/(1000~\rm s)$, we have 
\begin{eqnarray}
Y_{S,{\rm ID}} = 1.3\times 10^{-18} \times \frac{\Gamma_S}{10^{-3}\,{\rm Hz}} \times \left( \frac{1\,{\rm TeV}}{m_S}\right)^2 \nonumber\\\Longrightarrow~~ m_SY_S= 1.3 \times 10^{-15} \, {\rm GeV}\left.\right|_{m_S = 1\,{\rm TeV}}.
\end{eqnarray}
In spite of the large abundance, this falls about one order of magnitude short of the current best sensitivity \cite{Kawasaki:2017bqm}. In practice, the 
 sensitivity to singlet scalars is enhanced below their decay threshold to weak bosons, $m_S < 2 m_W$, where the 
decay width is set by the Yukawa coupling of $b$-quarks. Correspondingly, the same lifetime is achieved through a
parametrically larger value of $\theta$, which also translates into a larger abundance, and as a consequence, tighter BBN constraints. 

For $m_S \sim 100\,$GeV, the inverse decay processes is subject to significant uncertainties. 
In particular, it is not entirely clear what the true kinematic threshold is for $WW,ZZ,hh\to S$ production via the $2\to 1$ 
mechanism. For example, the thermal mass of longitudinal $W$'s is expected to be of order $m_{W,L} \sim (0.5-0.6) \times T$
at temperatures around the electroweak scale.
Thus, for $m_S \sim 100\,$GeV, it is difficult to determine for how long the $WW\to S$ process is kinematically accessible,
which renders predictions for the inverse decay processes very uncertain, and sensitive to the details of thermal 
physics near the electroweak cross over. 

Bearing these uncertainties in mind, and given our focus on the low $m_S$ range, to be conservative we will retain only the $2\to 2$ production channels in analyzing the constraints below. Note that an 
in-depth analysis would be required to achieve a higher precision calculation of the resonant and inverse decay production channels of 
$O(100 \,{\rm GeV})$ singlet scalar bosons.

\subsection{Thermalization of the $S$ sector with the SM}

The freeze-in abundance of  $S$ applies for mixing angles sufficiently small that the production rate remains below the Hubble rate,
\beq
\Gamma_{\rm prod} = n_i \left \langle  \sigma v\right\rangle_{12\to 3S} \lesssim H(T).
\eeq
Summing all the production channels, we find that the freeze-in relic abundance obtained with Eq.~(\ref{eq:sYdot}) is valid for
\beq
\theta \lesssim \theta_{\rm therm} \sim 10^{-6}.
\eeq
Larger mixing angles ensure complete thermalization with the SM bath before $S$ decouples and the relic abundance is simply given by the standard freeze-out paradigm. In this case, $Y_S$ is maintained at its relativistic equilibrium value
\beq
Y_{\rm eq} = \frac{45 \zeta (3)}{2\pi^2 g_\star (T)} \simeq \frac{0.28}{g_\star (T)},
\label{eq:Yeq}
\eeq
until $m_S$ becomes nonrelativistic, $T\lesssim m_S$, or the coannihilation rate becomes inefficient and $S$ decouples with its freeze-out abundance. Since $S$ interacts dominantly with heavy particles, $S$ remains relativistic while the coannihilating partners become nonrelativistic and the annihilation efficiency is exponentially lowered by the phase-space suppression of the other particles. Thus, $S$ freezes out according to Eq.~(\ref{eq:Yeq}) and the abundance only depends on the number of relativistic degrees of freedom $g_\star$ at the decoupling temperature. Above the QCD confinement scale $T_{\rm QCD}\sim 200 \MeV$, $g_\star$ varies by at most a factor of 2, in the range of $g_\star \sim 205/4 - 427/4$. A conservative estimate for the thermalized $S$ relic abundance is therefore
\beq
Y_S^{\rm f-o} \simeq \frac{1}{400}.
\label{eq:Yfo}
\eeq

\subsection{Validity of the Maxwell-Boltzmann approximation}

In the classic case of WIMP freeze-out, the decoupling temperature of the species is typically in the non-relativistic regime $T_{\rm decoupl} \sim m/20$. The statistical ensemble of particle energies is well described by the Maxwell-Boltzmann (MB) distribution, which allows for an analytical simplification of the phase-space integrals in the Boltzmann equation. In the freeze-in scenario considered here, this simplification is not necessarily justified and we must verify its validity. We derive in App.~\ref{sec:sYdot-fullstats} the analytical 3-dimensional expression to be numerically integrated for the $S$ abundance including the correct statistical distribution for all particles.

Instead of proceeding with the full treatment, we can verify the MB approximation with a simpler integration. Keeping the exact statistical distributions in the Boltzmann equation, we obtain~\cite{Edsjo:1997bg}
\begin{align}
s\dot{Y}&{=}\frac{1}{32\pi^4} \int ds\, p_{ij} \sqrt{s} \,\sigma \int_{\sqrt{s}}^\infty dE_+\int_{E_-^-}^{E_-^+} dE_-\, f_1 f_2 (1\pm f_3),  
\end{align}
with
\begin{align}
E_-^\pm &=\frac{|m_1^2-m_2^2|E_+}{s}\pm 2p_{ij}\sqrt{\frac{E_+^2-s}{s}}, 
\label{eq:sY_full}
\end{align}
where the initial energies were rewritten in terms of $E_+ = E_1 + E_2$, $E_- = |E_1 - E_2|$, $f_i$ is the Fermi-Dirac (FD) or Bose-Einstein (BE) distribution of species $i$ and the $+$ ($-$) is chosen for bosons (fermions) in the last term. The MB approximation~(\ref{eq:sYdot}) arises as an analytic solution in the MB limit $f_{1,2} = e^{-E_{1,2}/T}$ and $(1\pm f_3)\to 1$. We should stress that Eq.~(\ref{eq:sY_full}) is not mathematically correct as $E_3$ should have been included in the cross section phase-space integration over the end products. This integration is in general non-trivial and includes an additional angular dependence with $s$ (see App.~\ref{sec:sYdot-fullstats}). Nonetheless, we can use Eq.~(\ref{eq:sY_full}) as an estimate to the full result. To obtain the first correction beyond the MB approximation, we can expand
\beq
f_i = \frac{1}{e^{E_i/T}\pm 1} \simeq e^{-E_i/T} \left(1 \mp e^{-E_i/T} + \cdots \right).
\eeq
It is important to notice that the first order correction for the initial particles is equal to the MB limit of the $(1\pm f)$ term. At first order in $e^{-E_i/T}$, we have
\begin{align}
f_1 f_2 (1\pm f_3) &\simeq e^{-E_+/T} \left(1 + \kappa_1 e^{-(E_++E_-)/2T}   \right. \nonumber \\
&\quad  \left. +\kappa_2  e^{-(E_+-E_-)/2T} + \kappa_3 e^{-E_3/T}\right),
\label{eq:f_O1}
\end{align}
where $\kappa_i = \pm 1$, with $+$($-$) for bosons (fermions). As expected, the bosonic distribution enhances the overall yield, while the fermionic distribution decreases it. In principle, $E_3$ is a function of $\sqrt{s}$ and the angular kinematics for the final state particles. As mentioned, the $1\pm f_3$ factor should be included in the annihilation cross section, modifying $\sigma$. However, we know by conservation of energy that $E_1 + E_2 = E_3 + E_4$, with the following bounds on $E_3$
\beq
m_3 \leq E_3 \leq E_+.
\eeq
\begin{figure}[t]
\centering
  \includegraphics[width=.98\columnwidth]{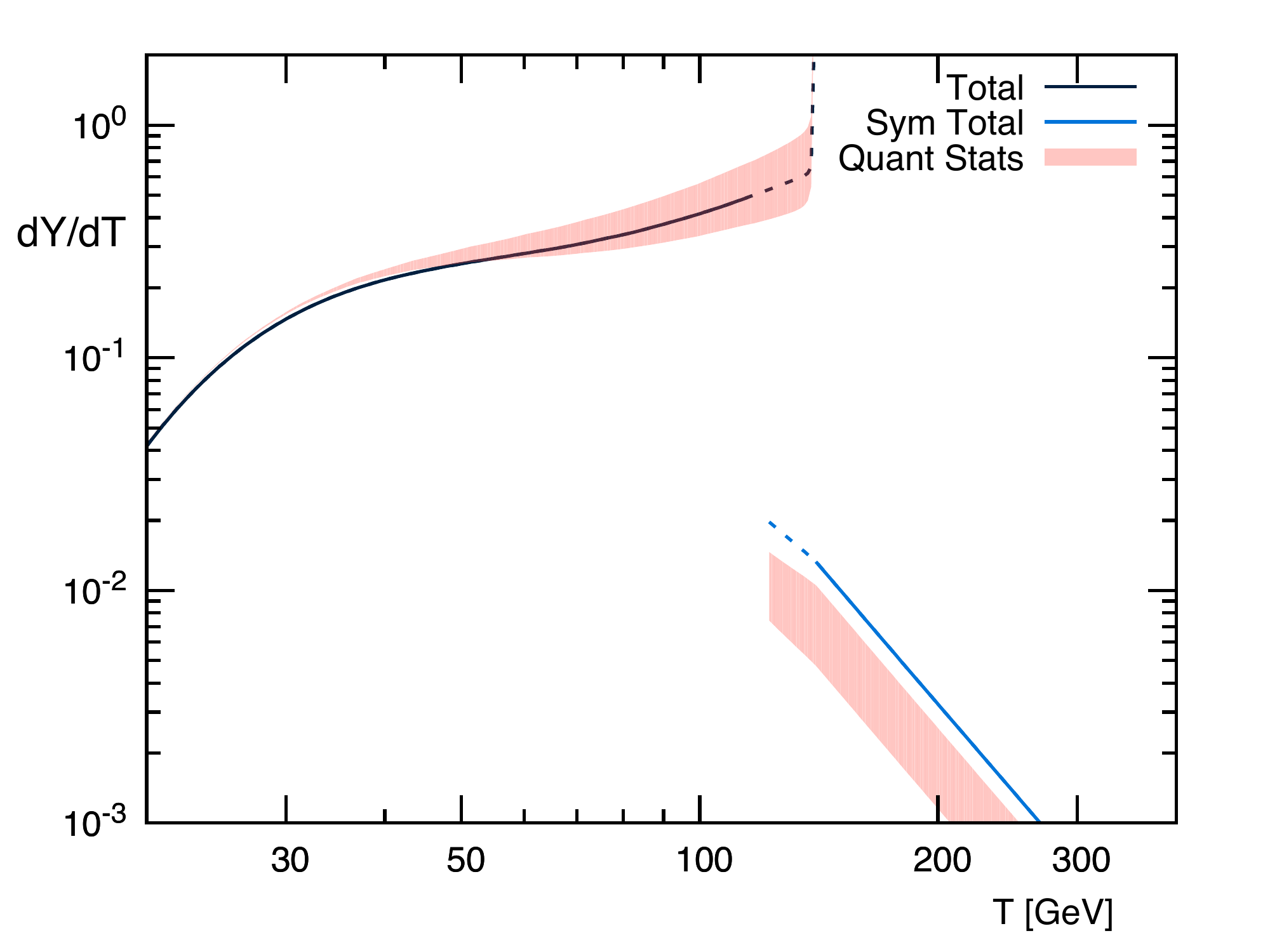}
\caption{The total $S$ emissivity as a function of temperature, including the estimated error range from adopting the MB approximation compared to  the correct emissivity with quantum distributions of particles 1, 2 and 3.}
\label{fig:Emissivity_QS}
\end{figure}
To estimate the range of possible yield values from the first correction to the full quantum distribution, we can integrate~Eq.~(\ref{eq:sY_full}) with~Eq.~(\ref{eq:f_O1}) for each of the $E_3$ extremum values. The potential spread in total emissivity in shown in Fig.~\ref{fig:Emissivity_QS} with an estimated range,
\beq
2.8\times 10^{11} \theta^2 < Y_S < 5.2 \times 10^{11} \theta^2. \label{MB}
\eeq
The total error for the MB approximation is thus expected to be within a factor of 2. The first order correction band in the symmetric phase lies completely below the MB value, because the top quark thermal masses in the QCD plasma dominate and suppress the available Fermi-Dirac distribution phase space.

\section{Cosmological constraints}

\label{sec:CosmoConst}

Having determined the freeze-in abundance for small mixing angles, we can now place the minimal set of bounds on the $S$ parameter space with $\theta \lesssim \theta_{\rm therm} \sim 10^{-6}$. We update and improve the cosmological constraints partially presented in both Ref.~\cite{Berger:2016vxi} and Ref.~\cite{Flacke:2016szy} and the final results for the low-$\theta$ parameter space are shown in Fig.~\ref{fig:ParamSpace}. The cosmological constraints that depend on $Y_S$ are discussed in the following subsections. Below $m_S \lesssim 5\keV$, the strongest constraint on $S$ comes from stellar energy loss~\cite{Hardy:2016kme} and even lighter scalars in the sub-eV range are constrained by $5^{\rm th}$ force experiments~\cite{Kapner:2006si,Decca:2007jq,Geraci:2008hb,Sushkov:2011zz} 
(as discussed in \cite{Piazza:2010ye}). We also show the projected sensitivity from the SHiP beam dump experiment~\cite{Alekhin:2015byh} and an order-of-magnitude estimate of supernova energy loss~\cite{Krnjaic:2015mbs} (which should be modified to account for in-medium effects \cite{Hardy:2016kme}). Above the pion threshold, we find strong sensitivity to the $S$ decay model. The colored exclusion regions presented in Fig~\ref{fig:ParamSpace} utilize the baseline decay model, which has a decay width which is larger than or equal to that in the spectator model. The baseline model therefore provides more conservative results due to the reduced abundance for a fixed lifetime.

\subsection{Diffuse X-ray background}

Many present-day satellites observe the galactic and extra-galactic photon spectrum in various wavelength bands and provide upper bounds on the luminosity of X-ray or Gamma-ray emission. These bounds apply for example to photons from decaying or annihilating dark matter. Below $m_S=2m_e$ and at small mixing angle, the lifetime of $S$ is longer than the age of the Universe, and therefore 
the model is constrained by these observations. 
In particular, Ref.~\cite{Essig:2013goa} derived the lifetime constraint on scalar dark matter particles decaying into 2 photons in the $4\keV < m_S < 10\GeV$ mass range assuming $\tau_S \gg \tau_{\rm universe}$. We directly rescale their constraint from the HEAO-1~\cite{Gruber:1999yr} and INTEGRAL~\cite{Bouchet:2008rp} satellites to obtain an exclusion band for $4\keV <m_S < 1\MeV$ with $10^{16}\sec\lesssim\tau_S \lesssim10^{22}\sec$ displayed as \textit{X-Ray} in Fig.~\ref{fig:ParamSpace}.

\subsection{CMB Anisotropies}

Precision measurements of the temperature and polarization anisotropies in the CMB by the WMAP~\cite{Hinshaw:2012aka} and Planck~\cite{Ade:2015xua} satellites provide strong constraints on energy injection that can ionize cosmic neutral hydrogen after recombination~\cite{Chen:2003gz,Pierpaoli:2003rz,Padmanabhan:2005es,Zhang:2007zzh,Finkbeiner:2011dx,Slatyer:2016qyl}. The raised ionization fraction at lower redshifts allows for delayed photon interactions, which modifies the visibility function that weighs the probability of last scattering for a given CMB photon at a specific time. This effectively damps the high-$l$ tail of the TT power spectrum and increases the low-$l$ E-mode polarization~\cite{Chen:2003gz,Pierpaoli:2003rz}.

\begin{figure}[t]
\centering
 \includegraphics[width= 0.98\columnwidth]{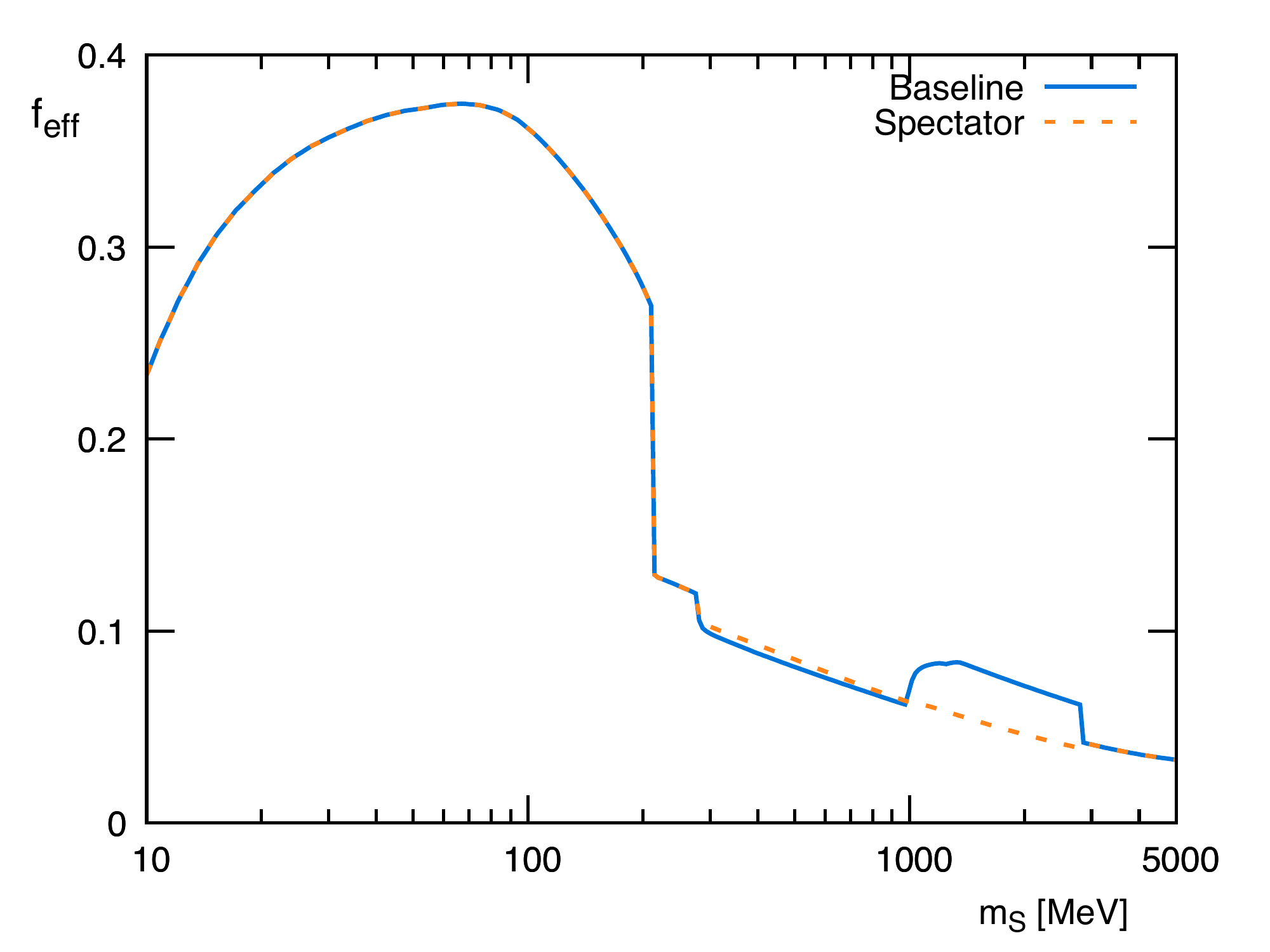}
  \caption{Effective fraction of energy deposited leading to ionization of the cosmic plasma at $z = 300$ for $\Gamma_S = 10^{14}$~s, for the baseline and spectator decay models.}
\label{fig:feff}
\end{figure}

At redshift $z_{dep}$, the efficiency of energy deposition in the cosmic plasma by an energetic electron-positron pair or photons injected at an earlier redshift $z_{inj} > z_{dep}$ has been determined in Ref.~\cite{Slatyer:2012yq} and updated in Ref.~\cite{Slatyer:2015kla}. This update provides the energy fractions that go into ionization, excitations, heating and emission of low-energy photons. Given the process-dependent and $z$-dependent ionization efficiency, comparing the modified power spectra to the CMB data is computationally intensive. In practice, principal component analysis of modified recombination histories shows that a decaying particle is well described by a constant deposition efficiency taken at $z_{dep}=300$~\cite{Slatyer:2015kla,Poulin:2016anj}. We can then simply utilize the derived constraints for decaying particles in Ref.~\cite{Fradette:2014sza} and translate to the current model with
\beq
\zeta = f_{\rm eff} \frac{m_SY_S s_0}{m_p n_{b,0}},
\eeq
where $f_{\rm eff}= f(z=300)$ is the ratio of energy absorbed leading to ionization over energy emitted at $z_{\rm dep}= 300$. In the mass range where $ S \to \mu^+ \mu^-$ is the main decay channel, we solve for $f_{\rm eff}$ by integrating over the electron from muon decay, which decreases the ionization efficiency by a factor of 3 due to neutrinos radiating away energy. We repeat the procedure for the decay chains $S \to \pi^+ \pi^- \to \mu^+ \mu^- \nu \nu$, $S\to \pi^0 \pi^0 \to \gamma \gamma \gamma \gamma$ and find that it is well approximated by evaluating the decay products at their average energy from the decay. We evaluate the efficiency of kaons by weighting the main branching ratios and the decay products by their average energy, percolating down to their final $e^\pm-\gamma-\nu$ spectra. Above the di-charm threshold, the light quarks, charm quark and gluon all have similar deposition efficiencies that lie somewhere between those of electrons and muons~\cite{Slatyer:2016qyl}. In general, for $m_S \lesssim 10\GeV$, the efficiency tends to approach the muonic case~\cite{Cline:2013fm}. We adopt the same ionization efficiency as muons for conservative results. The overall $f_{\rm eff}$ for $S$ with $\tau_S = 10^{14}\sec$ is shown in Fig.~\ref{fig:feff} for both the baseline and spectator decay models.

\begin{figure}[t]
\centering
 \includegraphics[width= 0.98\columnwidth]{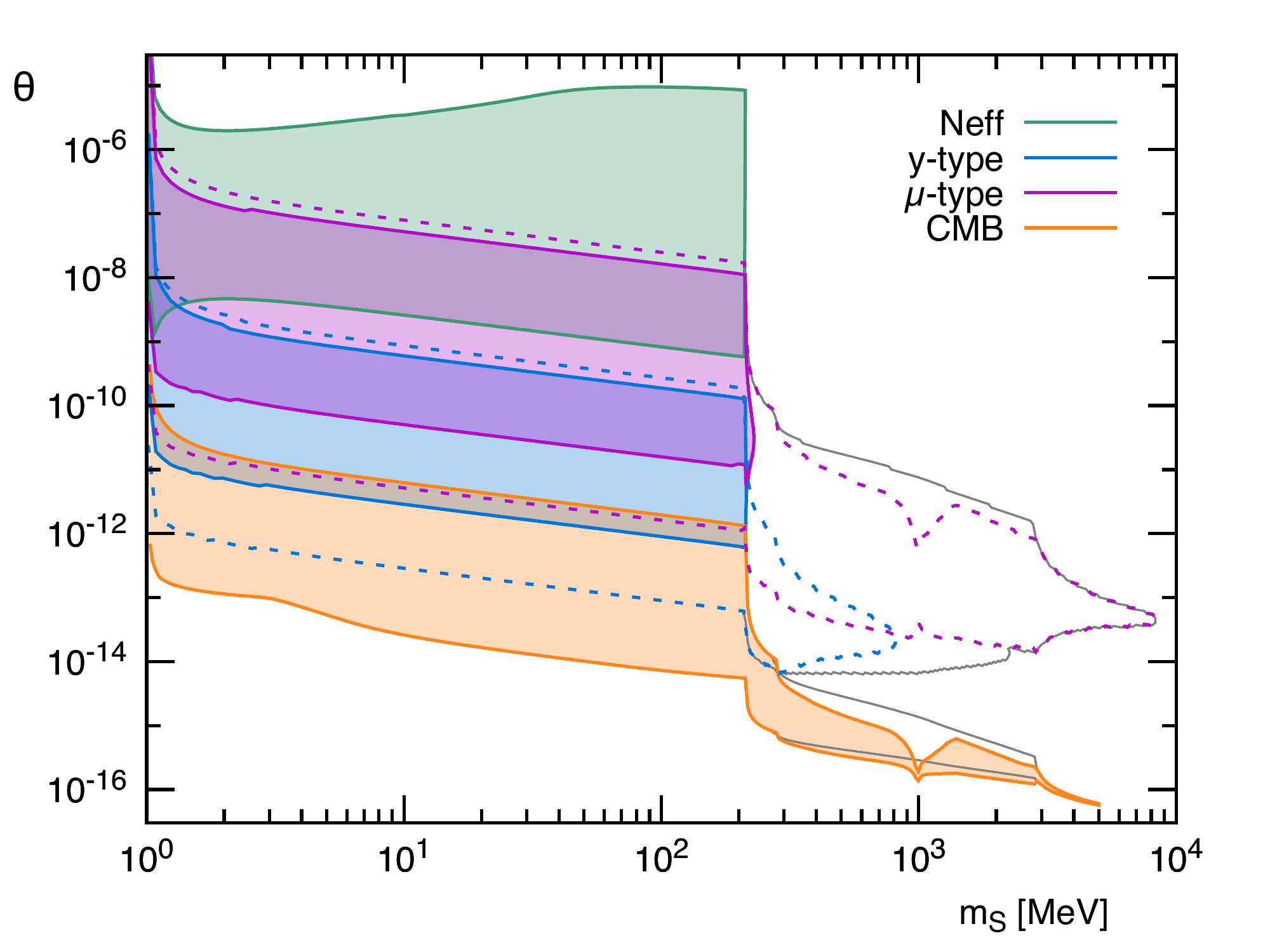}
  \caption{Detailed cosmological constraints on $S$ in the MeV-GeV mass range, excluding BBN (see Fig.~\ref{fig:BBN}). The solid lines and shaded areas represent the parameters excluded in the baseline decay model. Dashed lines refer to the projected sensitivity to spectral distortions of a PIXIE-like detector~\cite{Kogut:2011xw} and the thin gray line exhibits the would-be PIXIE sensitivity in the spectator $S$ decay model.}
\label{fig:CMB-SpecDist}
\end{figure}

We will not extrapolate the CMB constraints down to lifetimes $\tau_S < 10^{13}$~s because $f_{\rm eff}$ is not numerically stable for decays before recombination~\cite{Slatyer:2015kla} and the on-the-spot approximation at $z_{\rm dep}=300$ fails to represent the correct physics for short lifetimes~\cite{Poulin:2016anj}. The excluded band is shown in orange within Fig.~\ref{fig:CMB-SpecDist} for the baseline model, with the would-be exclusion region for the spectator decay model delimited by a thin gray line. 

\subsection{CMB Spectral distortions}

\label{sec:SpectDist}

While energy injection after recombination may be observed in the CMB as variations in the anisotropies, earlier energy injection can induce spectral distortion of the blackbody distribution~(see Ref.~\cite{Chluba:2011hw} for a recent review), and can be used as probes of decaying particles~\cite{Chluba:2013wsa,Chluba:2013pya}.

Cosmological thermalization is very efficient at arbitrarily early times and the photon plasma becomes susceptible to incomplete re-equilibration of its spectrum for energy injected at $z \lesssim z_\mu \simeq 2 \times 10 ^6$. The CMB photons are still efficient at redistributing their energy across the energy spectrum, but double-Compton scattering and Bremsstrahlung interactions that adjust the number of photons become inefficient. The bath thus develops a non-zero chemical potential in its high-energy tail resulting in the $\mu$-distortion. At lower redshifts, $z \lesssim z_{\mu y} \simeq 5 \times 10^4$, Compton scattering between electrons and photons fails to maintain both species at a common temperature. The photon bath inherits a reduced temperature at low energies while high frequencies receive a relative gain in temperature, a phenomenon called the Compton $y$-distortion~\cite{Chluba:2011hw}.

Distortion due to arbitrary energy injection can be approximated by~\cite{Chluba:2016bvg}
\begin{align}
y &= \frac{1}{4} \int_{z_{\rm rec}}^{z_{\mu y}}\frac{d (Q/\rho_\gamma)}{dz'} dz',  \\
\mu &= 1.401 \int_{z_{\mu y}}^\infty e^{-\left(\frac{z'}{z_\mu}\right)^{5/2}} \frac{d (Q/\rho_\gamma)}{dz'} dz',
\end{align}
where $z_{\rm rec} = 1000$ and the normalized injected electromagnetic energy is
\beq
\frac{d (Q/\rho_\gamma)}{dz'}  =\frac{1}{\rho_\gamma}\frac{d E}{dt dV} Br_{\rm em}\frac{1}{H(1+z')}.
\eeq
In this expression, $\frac{d E}{dt dV}$ is the total energy injected and $Br_{\rm em}$ is the branching ratio to electromagnetic end products. In a radiation-dominated Universe, the $y$-distortion can be evaluated analytically,
\beq
y\simeq \frac{\sqrt{\pi}}{8} \frac{Y_{S} m_{S} s_0}{\rho_{\gamma 0}\sqrt{\Gamma_S t_0}}\; Br_{\rm em}\;\mathcal{I}(\Gamma),
\eeq
with the current entropy density $s_0 = 2891 \;{\rm cm}^{-3}$, the current photon energy density $\rho_{\gamma 0} = 0.26 \;{\rm eV}\; {\rm cm}^{-3}$, a time normalization of $t_0 = 2.4 \times 10^{19} \sec$ and where the integral $\mathcal{I}(\Gamma)$ is defined as
\begin{align}
\mathcal{I}(\Gamma) &= \frac{2}{\sqrt{\pi}} \int_{\frac{\Gamma t_0}{z_{\mu y}^2}}^{\frac{\Gamma t_0}{z_{\rm rec}^2}} e^{-\xi} \sqrt{\xi} \; d\xi \\
 &\to \left\{ 
\begin{array}{rl} 1, & {\rm if}\; 10^{-13} \lesssim \Gamma \times {\rm sec} \lesssim  10^{-10}, \\0, &  {\rm if}\;\Gamma \ll 10^{-13}\; {\rm sec}^{-1}\;\;\; {\rm or}\;\;\; \Gamma \gg 10^{-10}\; {\rm s}^{-1}.\end{array}\right. \nonumber
\end{align}
The measured bounds from COBE/FIRAS~\cite{Fixsen:1996nj} and the projected sensitivity from a PIXIE-like detector~\cite{Kogut:2011xw} are 
\begin{align}
\mbox{COBE/FIRAS:} &&|y| &\leq 1.5 \times 10^{-5}&|\mu| &\leq 9 \times 10^{-5},\\
\mbox{PIXIE:} &&|y| &\leq 2 \times 10^{-9}&|\mu| &\leq 1 \times 10^{-8}.
\end{align}

\begin{figure}[t]
\centering
 \includegraphics[width= 0.98\columnwidth]{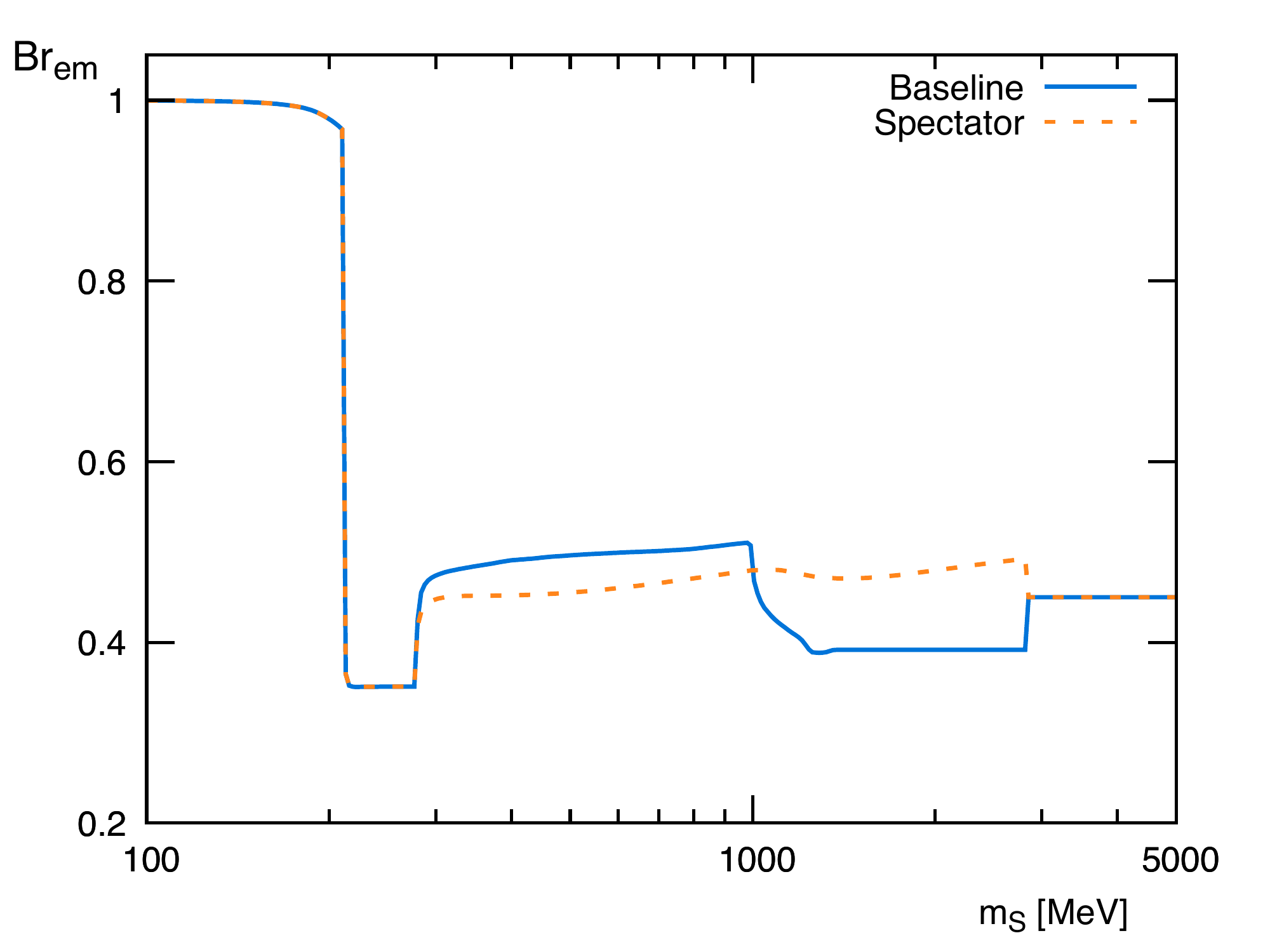}
  \caption{Fraction of $S$ rest energy decaying into electromagnetic energy as a function of its mass for the baseline and spectator decay models.}
\label{fig:BrEM}
\end{figure}

We approximate the electromagnetic branching ratio by weighting the average energy carried by end products from initial decays at rest with their respective branching ratios from $S$. Since the averaged energy carried away by the electron in a muon decay is $\langle E_e \rangle / m_\mu = 0.35$~\cite{Pospelov:2010cw}, we have $Br_{\rm em}^{S\to \mu^+ \mu^-} = 0.35$. The electromagnetic 
fractions for heavier decay products can be found from the corresponding fractions of their lighter 
decay products. We find the following electromagnetic energy injection ratios
\begin{align}
E_{\rm em}^{\mu^\pm} &\simeq 0.35\times E_{\mu^\pm}, 
&E_{\rm em}^{\pi^\pm} &\simeq 0.27 \times E_{\pi^\pm},\\ E_{\rm em}^{\pi^0} &= 1.00 \times E_{\pi^0}, 
&E_{\rm em}^{K^\pm} &\simeq 0.29\times E_{K^\pm}, \\
E_{\rm em}^{K^0_S} &\simeq 0.49 \times E_{K^0_S}, & E_{\rm em}^{K^0_L} &\simeq 0.48 \times E_{K^0_L},
\end{align}
where we neglected the kaon decay channels that contribute less than 10\% of the kaon decay width. The total electromagnetic branching ratios for the $S$ decay channels are
\begin{align}
Br_{\rm em}^{S\to \mu^+\mu^-} &= 0.35, &Br_{\rm em}^{S\to \pi\pi} &= 0.51, \\
Br_{\rm em}^{S\to KK} &= 0.39, &Br_{\rm em}^{S\to q\bar{q},gg} &= 0.45,
\end{align}
where 2/3 (1/3) of pions are charged (neutral), 1/2 (1/4+1/4) of kaons are charged (neutral short + long) and we assume the electromagnetic yield of high energy quarks and gluons of 0.45 is maintained for $c$-quarks. The total $Br_{\rm em}$ as a function of $m_S$ is shown for the baseline and the spectator decay models in Fig.~\ref{fig:BrEM}. The excluded regions from COBE/FIRAS are shown in Fig.~\ref{fig:CMB-SpecDist}, with a robust conservative overlap between all probes in the $1\MeV < m_S < 2 m_\mu$ mass range. A PIXIE-like detector would not change the constraints in the $m_S < 2m_\mu$ mass range, but has the potential to increase the sensitivity range to $m_S \lesssim 8 \GeV$, with a sensitivity band that somewhat depends on the $S$ decay model.

\subsection{Relativistic degrees of freedom (\boldmath{$N_{eff}$)}}

\label{sec:Neff}

The total relativistic energy density in the Universe at decoupling is well constrained by the CMB. The temperature of the photon bath determines its contribution to the total radiation energy density. Any additional component is parametrized in $N_{\rm eff}$, the effective number of neutrinos with a temperature $T_\nu = (4/11)^{1/3} T_\gamma$. The Planck collaboration measurement of $N_{\rm eff} = 3.04 \pm 0.33$~\cite{Ade:2015xua} at $2\sigma$ is in agreement with the SM predicted value of 3.046~\cite{Mangano:2005cc}.

The constraints on early injection $\tau_S < 1 \sec$ of $S\to \{e^+ e^-,\gamma \gamma\}$ or $S \to \mu^+ \mu^-$ have been derived in Ref.~\cite{Fradette:2017sdd} including neutrino decoupling effects, which we apply using the freeze-out abundance from Eq.~(\ref{eq:Yfo}) in the mass range $10\MeV \lesssim m_S < 2 m_\mu$. For lower masses, the $S$ lifetime at the border of the exclusion band is longer than $\tau_S = 100$~s, the maximal range derived in Ref.~\cite{Fradette:2017sdd}. For longer lifetimes, we simply compare the energy density of the $S$ sector with the SM energy densities in the neutrino and EM baths prior to the energy release (at $t = x \tau_S$) and assume an instantaneous decay. $N_{\rm eff}$ can then be estimated by comparing the relative energy density in the neutrino and EM baths
\beq
N_{\rm eff} = \frac{8}{7} \left( \frac{11}{4}\right)^{4/3} \frac{\rho^0_\nu+ \rho_\nu^S}{\rho^0_\gamma+\rho^S_\gamma},
\eeq
where $\rho_\gamma^S = Br_{\rm em} \rho_S (t = x \tau_S)$, $\rho_\nu^S = (1-Br_{\rm em}) \rho_S (t = x \tau_S)$ and the electromagnetic/neutrino energy partition of the end products $Br_{\rm em}$ is taken from Fig.~\ref{fig:BrEM}. We choose $x=1/10$ to match the constraints for $\tau_S < 100$~s. This choice of $x=1/10$ is conservative. The nonrelativistic $S$ energy density decreases less with time than the relativistic SM energy density and larger values of $x$ would yield stronger constraints. Using the $2\sigma$ range for $N_{\rm eff} = 3.04 \pm 0.33$, we find the constraints labelled $N_{\rm eff}$ in Figs.~\ref{fig:ParamSpace} and~\ref{fig:CMB-SpecDist}. 

\subsection{Big Bang Nucleosynthesis}

The synthesis of light nuclei during BBN is well understood, with the final abundance of stable light elements in good agreement with predicted values (see for example~\cite{Cyburt:2015mya,Coc:2017pxv,Pitrou:2018cgg} for recent overviews and discussions of the discrepancy with ${}^7$Li). The concordance between predicted and observed abundances of \Hefour, $^3$He and D can be used to constrain electromagnetic and hadronic energy injection~(see Refs.~\cite{Jedamzik:2009uy,Pospelov:2010hj,Kawasaki:2017bqm} for reviews).

The impact of decaying particles in the BBN era depends on the ability of the decay products to efficiently interact with light nuclei, which varies as the BBN reaction network evolves and the universe cools down. Early mesons decays were thoroughly discussed in Ref.~\cite{Fradette:2017sdd}, effectively increasing the $n/p$ ratio before they freeze out, thus raising the $^4$He yield above its observational limit. After most neutrons have converted to \Hefour, the negatively charged mesons, $\pi^-$ and $K^-$ can dissociate the copious \Hefour, producing lighter \Hethree, T, D, $n$ and $p$ that are fed back into the reaction network~\cite{Pospelov:2010cw}. This mechanism was suggested to decrease the $^7$Li prediction by reducing the amount of ${}^7{\rm Be}$ and resolving the discrepancy with observations~\cite{Pospelov:2010cw}, but incidentally also raising the $D/H$ ratio above $3\times 10^{-5}$ from inefficient $D$ burning, which is now excluded by observations~\cite{Cooke:2013cba,Cooke:2016rky}. Beyond $\tau_S \gtrsim 10^4$~s, the mesonic interaction rate with ambient nuclei is suppressed below the decay rate by the dilution due to expansion. The mesons instead have enough survival time to decay away in a shower of electromagnetic energy. Photodissociation of nuclei becomes efficient when photons have cascaded down below the $e^+e^-$ thermal pair creation energy $E_{\rm th} = m_e^2/22T$. These $\gamma$-rays can photodissociate D with a binding energy of $E_{\rm D}^{\rm bind} = 2.2\MeV$ at $t\sim 5\times 10^4$~s and similarly for $^4$He with $E_{{}^4{\rm He}}^{\rm bind} = 19.8\MeV$ at $t\sim 4\times 10^6$~s~\cite{Jedamzik:2006xz}.

\begin{figure}[t]
\centering
 \includegraphics[width= 0.98\columnwidth]{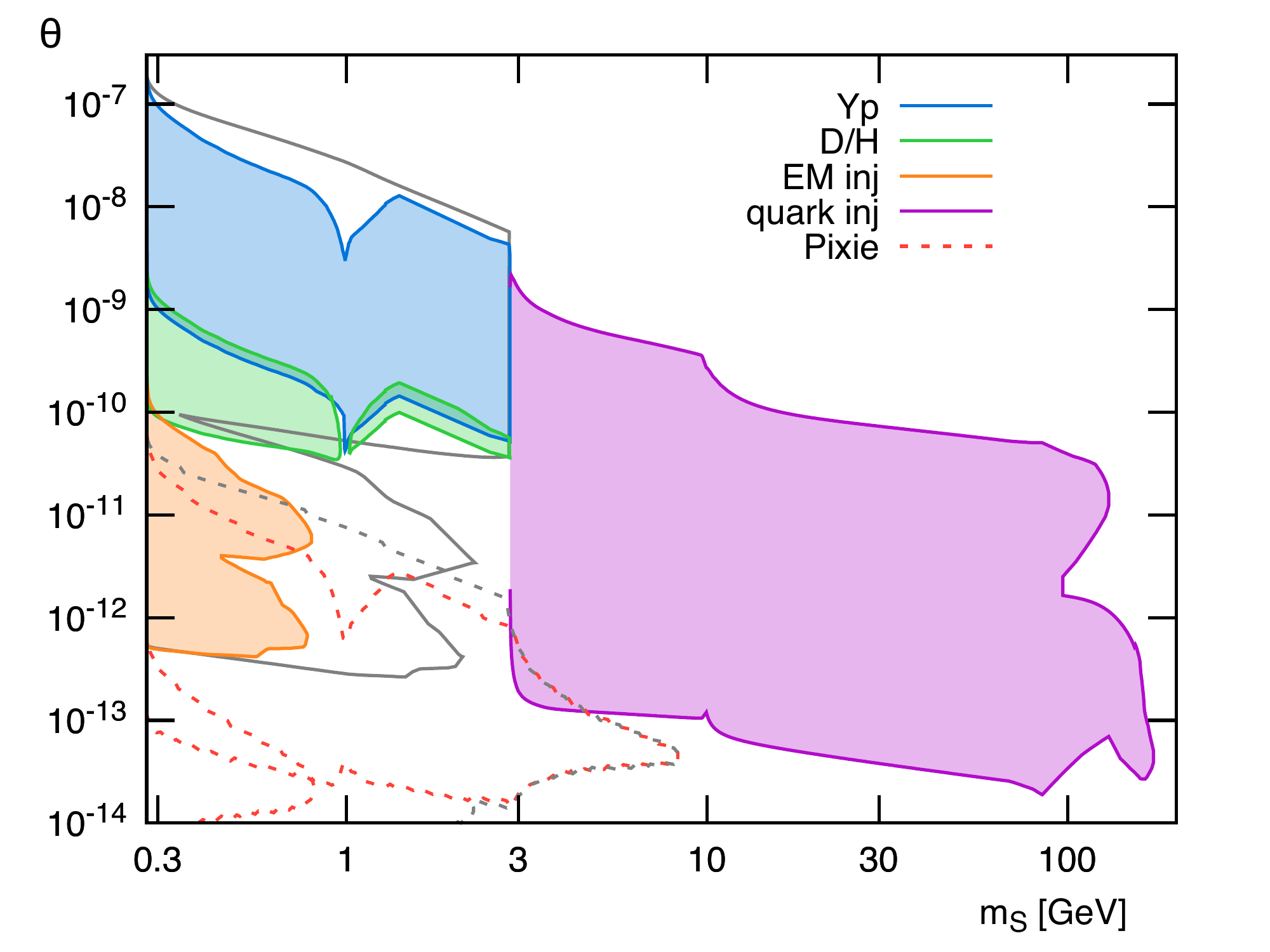}
  \caption{BBN constraints above the di-pion threshold. Solid lines and shaded areas are ruled out if $S$ follows the baseline decay model. Thin gray lines show the region for the spectator decay model instead. For comparison, dashed lines represent the future spectral distortion sensitivity from a PIXIE-like detector~\cite{Kogut:2011xw}.}
\label{fig:BBN}
\end{figure}

We implement mesonic decays of $S\to \pi \pi$ and $S \to K K$ (charged and neutral) by weighting the freeze-in abundance by their respective ratios. We can then apply the constraint on early decays from Ref.~\cite{Fradette:2017sdd}, constraining the $S$ parameter space by an overproduction of $^4$He shown in blue in Fig.~\ref{fig:BBN}. Moreover, we use the $D/H<3\times 10^{-5}$ limit from Ref.~\cite{Pospelov:2010cw}, utilizing their stopped pion and kaon analysis for conservative results, displayed as the orange region on the $S$ parameter space. Longer lifetimes with electromagnetic shower constraints are shown in green. We weight the $S$ abundance by the electromagnetic branching ratio from Fig.~\ref{fig:BrEM} and compare the value with the electromagnetic injection limit from Ref.~\cite{Kawasaki:2017bqm}. The upper protruding band is from a decreased D/H ratio while the lower green region is from an increased \Hethree/D ratio. We have assumed 100\% decays to kaons in the uncertain region, from $m_S \sim 1.4\GeV$ to the di-charm threshold. Note that the BBN constraint on $S$ from mesonic decays has large uncertainties due to the unknown decay spectrum. This is illustrated by the rather different exclusion region if the spectator decay model is assumed instead, as shown by the thin gray line in Fig.~\ref{fig:BBN}.

Above the di-charm threshold, the $S$ decay products are well understood and the model allows for more accurate predictions. BBN is sensitive to the total energy injected via quark pairs, with a negligible dependence on the quark flavor~\cite{Kawasaki:2017bqm}. The sensitivity is mostly dominated by the number of hadrons produced in the hadronic shower, with a $m_S^{0.3}$ dependence on the total energy input (assuming a nonrelativistic $S$). Constraints from quark injection therefore increase with a smaller $m_S$~\cite{Kawasaki:2017bqm}. We use the $b\bar{b}$ limits from a 30 GeV initial particle from Ref.~\cite{Kawasaki:2017bqm} for $c\bar{c}$ and $b\bar{b}$ injection, both rescaled by their respective branching ratio. The resulting exclusion is shown in red in Fig.~\ref{fig:BBN}, with the upper $\theta$ range constrained by $^4$He overabundance and the lower $\theta$ values excluded by a D overproduction. 

Electromagnetic injection below the pion threshold will also affect the BBN network, through decays to muons and electrons with a nontrivial dependence on the $S$ mass~\cite{Berger:2016vxi}. However, the $S$ decay rate is much weaker and requires $\theta$ values close to $S$ thermalization in order to decay during the active BBN epoch. In similar regions of parameter space, there are also significant constraints from the limits on $N_{\rm eff}$, derived in section~\ref{sec:Neff}, along with the spectral distortion results from section~\ref{sec:SpectDist}. These already exclude the parameter space with $\tau_S \sim 0.1-10^{13}$~s for $2 m_e < m_S < 2m_\mu$. We therefore do not need to perform a detailed analysis of the BBN constraints below the di-pion threshold and, by the same token, the potential solution to the $^7$Li problem tagged in Ref~\cite{Berger:2016vxi} is also ruled out. 

\section{Discussion and Conclusions}

\label{sec:Discuss}

Interacting light scalar bosons are usually associated with a familar technical naturalness problem, 
$m_{\rm scalar} \sim ({\rm couplings}) \times \Lambda_{\rm UV}$, resulting in the dilemma 
of justifying fine-tuning. The super-renormalizable portal offers an exception:
the mass of the scalar can be natural both at tree and loop level, as long as $m_S$ is larger than the trilinear coupling $A$,
or in terms of the mixing angle, $m_S > \theta \times (100\,{\rm GeV})$. Only a very few particle physics experiments 
(notably flavour changing decays $b\to s S,\, s\to d S$ \cite{Batell:2009jf}) and the 5th force experiments \cite{Piazza:2010ye}  
can probe the parameter space with natural values of couplings. In contrast, cosmology within the $\Lambda$CDM framework 
provides a sensitive probe of this model, with couplings reaching down to $\theta \sim 10^{-16}$, 
which in terms of the coupling to electron translates to $g_{Se} = \theta \times (m_e/v) \propto 10^{-21}$. (Incidentally, 
gravitons have approximately the same coupling strength to nonrelativistic electrons.) In this paper, we have attempted to take full 
advantage of the available cosmological sensitivity, by improving the calculations of metastable $S$ abundance produced through the 
unique super-renormalizable portal the SM has to offer: $ASH^\dagger H$. 

We have revisited the calculation of freeze-in production of the Higgs portal scalar and shown that electroweak gauge bosons and the $t-b$ quarks have non-negligible contributions to the total yield. %
We find the largest yield from top quark coalescence with a soft $S$ emission, regulated by the finite top quark width. Even for $m_S \ll m_h$, the favoured interactions with heavy particles push the bulk of the production just below the electroweak phase transition, in the $10\GeV \leq T \leq T_c$ temperature range. Improvements in the precision of the calculation require the full finite-temperature quantum field theory machinery that models the phase transition and includes plasma screening effects. This incremental effort is substantial and not necessarily justified for new physics searches, especially because we estimate the current level of precision to be a factor of $\sim 2$.

Control of the abundance calculation has allowed us to provide a full parameter space scan of the low mixing angle constraints, from sub-eV masses to 100\GeV. The cosmological constraints that depend on the primordial $S$ abundance appear for $m_S > 10\keV$. The Higgs portal scalar has well-defined decay products below the pion threshold. It allows for rigorous constraints in the $10\keV < m_S < 2 m_\mu$ range that fully cover 6 to 10 orders of magnitude in mixing angle. The low $\theta$ boundary is probed by the late decays seen in the diffuse X-Ray background and the CMB. In both cases, the lifetimes exceed the sensitive injection time window. This means that the number of decaying particles scales as $dN/dt \propto Y_S\Gamma_S \propto \theta^4$.  We note at this point that the dependence on $({\rm coupling})^4$ is a familiar one: it appears 
in any beam dump experiment searching for a decay within a detector in the limit where the decay length is much larger than the 
size of the experiment. It also appears in the ``cosmological beam dump" constraints considered here, 
for the case of extremely long-lived particles. 
The change in $S$ abundance by a factor of 2 would change the $\theta$ sensitivity by a factor of $2^{1/4}$, which would not 
be visible on the scale of the final log-log plot.  
For this reason, improvements in the calculation of $Y_S$ would provide a minimal gain in precision and our approximate framework with near-vacuum cross sections is justified, at least for $m_S < 2 m_\mu$. Similarly, the upper boundary for large $\theta$ is set by the freeze-out relic abundance, rather than the freeze-in abundance. Early cosmological energy injection has a logarithmic dependence on the $S$ primordial metastable abundance~\cite{Fradette:2017sdd}. This log dependence renders the exclusion region robust to variations of a few in $Y_S^{\rm f-o}$, confirming the accuracy of our conservative estimate, Eq.~(\ref{eq:Yfo}), without the need for a relativistic freeze-out computation.

Above the pion threshold, in the $2m_\pi < m_S \lesssim 2 m_c$ mass range, we have derived the CMB anisotropy, spectral distortion (plus the forecast for a PIXIE-like detector), and BBN constraints for two different decay models, with significant differences in the exclusion bands. Especially near the kaon resonance, $m_S \sim 1 \GeV$, the sensitivity of each probe can vary by an order of magnitude in $\theta$. Given this poorly defined feature, we stress that improvements in the determination of the $S$ mesonic decay width are needed before finite-temperature abundance calculations are warranted to improve the accuracy of cosmological probes.

For higher masses, where $S$ predominantly decays to quarks, the large $\theta$ sensitivity boundary diverges sharply for short lifetimes in the $m_S Y_S$ vs $\tau_S$ plane~\cite{Kawasaki:2017bqm}. Thus the $\theta$ constraint is insensitive to minor changes in $Y_S$. For $\tau_S > 10^4$~s, however, the required energy stored in the dark sector, $m_S Y_S$, is almost flat as a function of increasing lifetimes. Changes in the freeze-in yield by a factor of 2 can thus be compensated by a factor of $2^{1/4}$ in the mixing angle.
We therefore conclude that cosmological constraints on $S$ are robust to variations of $Y_S$ by a factor of a few, except where $S$ decays primarily to mesons, where the determination of the decay width results in a larger uncertainty.

The derived constraints are relatively conservative with respect to the structure of the dark sector provided that a quartic interaction $\lambda_S S^2 H^\dagger H$ does not thermalize the dark sector. The only additional requirement 
for the existence of these constraints is that $S$ decays visibly and not into some 
additional stable  dark state. In this case, the freeze-in production mechanism provides the minimal metastable abundance $S$ can have for a given lifetime. Any additional interactions with other states will increase its population, until it thermalizes with the SM. Large self-interactions can potentially dilute $Y_S$ before it decays, which would reduce the $S$ abundance by a factor of $\ln (a_{\tau_S}/a_{\rm f-i})$~\cite{Carlson:1992fn}. Given that the freeze-in relic abundance is set at $T_{\rm f-i} \sim 10 \GeV$, this is a negligible factor of a few for early decays probed by BBN or $N_{\rm eff}$, but can potentially be more than an order of magnitude in $Y_S$ for lifetimes relevant to the CMB and X-ray measurements. Large self-interactions could therefore reduce the low $\theta$ sensitivity by a factor of a few.

\section*{Acknowledgments}

We thank Chien-Yi Chen, Robert Lasenby, Subir Sarkar and Roman Zwicky for helpful discussions. The work of MP and AR is supported in part by NSERC, Canada, and research at the Perimeter Institute is supported in part by the Government of Canada through NSERC and by the Province of Ontario through MEDT. JP is supported by the New Frontiers program of the Austrian Academy of Sciences.

\appendix

\begin{widetext}

\section{Scalar decays below the electron threshold}

As summarized in Section~IIA, below the electron mass threshold a Higgs-like particle decays to 2 photons at leading order through a loop of heavy particles~\cite{Djouadi:2005gi},
\begin{align}
\Gamma (S\to \gamma \gamma) &= \frac{\theta^2 \alpha^2 m_S^3}{256 \pi^3 v^2} \left| C \right|^2, 
\end{align}
where
\begin{align}
C &= \sum_f N_c Q_f^2 A_f (\tau_f) + A_W (\tau_W),
\end{align}
in terms of loop functions $A_f$ and $A_W$ which are functions of $\tau_X = m_S^2/4m_X^2$. Note that the $W$ contribution has the opposite interference sign, and the functions are given by
\begin{align}
A_f (\tau) &= \frac{2}{\tau^2} \left[\tau + (\tau -1 ) f(\tau)\right]  \xrightarrow[\tau \to 0]{} 4/3, \\
A_W (\tau) &= \frac{-1}{\tau^2}\left[ 2\tau^2 + 3\tau + 3(2\tau -1) f(\tau)\right] \xrightarrow[\tau \to 0]{} -7,
\end{align}
with
\beq
f(\tau) = \left\{ 
\begin{array}{lr}
\arcsin^2 \sqrt{\tau} & \tau \leq 1,\\
-\frac{1}{4} \left[ \ln \frac{1+\sqrt{1-\tau^{-1}}}{1-\sqrt{1-\tau^{-1}}}-i\pi\right]^2 & \tau>1.
\end{array} \right. 
\eeq
In this prescription, the inclusion of light quarks $u$, $d$ and $s$ is rather ambiguous. Their masses are not physical rest masses since  $m_{u,d,s} \ll \Lambda_{\rm QCD} \simeq 350 \MeV$. Instead, they are determined as chiral symmetry breaking variables in QCD which generate the non-zero mass of the Goldstone bosons of the approximate symmetry, the pions, kaons and eta mesons~\cite{Pich:1995bw}. A more appropriate interpretation of their contribution to $S\to \gamma \gamma$ is through virtual loops of pions and kaons~\cite{Leutwyler:1989tn}. Adding the contributions from all SM particles, for $m_S \ll 2m_e$ we find
\beq
C = \left\{ 
\begin{array}{llr}
11/3 & \simeq 3.67&\qquad\mbox{for 0 + 6 quarks}  \\
989/522 & \simeq 1.89&\mbox{for 2 + 4 quarks} \\
50/27 & \simeq 1.85&\mbox{for 3 + 3 quarks} \\
1 & = 1&\mbox{for 0 + 3 quarks}
\end{array} \right. 
\eeq
where different scenarios of ($a$ light) + ($b$ heavy) quarks are shown. For the case of 2(3) light flavours 
the pion (pion and kaon) loops are taken into account, while for 0 light flavours they are neglected. 
The true physical value should be close to the $3+3$ or $2+4$ scenarios. 

\section{Higher order corrections to the thermal mixing angle}

As discussed in Section~3, for small $m_S$ the mixing angle $\theta$ scales quite differently at zero temperature and at temperatures close to the 
phase transition. From Eq. (\ref{mixingT=0}), we find $\theta \propto A/(\lambda_H v)$,  while at finite temperature $\theta_{\rm eff}(T)$, given by Eq.~(\ref{thetaTeff}), scales quite differently as $\theta_{\rm eff}(T) \propto Av(T)/({\rm couplings}\times T^2)$, particularly as $v(T)\rightarrow 0$ at the phase transition. The resolution is that there are additional diagrams which contribute to these vertices at higher order in perturbation theory and survive in the limit of $v(T)\to 0$. Example diagrams in the higher order expansion are shown in Fig.~\ref{fig:ZZS_sym}.

The tree level coupling between $Z$ and $S$ is proportional to $g^2Av^2(T)/m_h^2(T) $, and vanishes in the limit 
$v(T)\to 0$. However, it is clear that the thermal average of the 
full Higgs including vev and fluctuations, $\langle (v+h)^2(T) \rangle$, does not vanish.
This follows from the first thermal correction, with one thermal loop of Goldstone and Higgs bosons. 
Taking the soft $S$ limit so that the Matsubara sum of the Goldstone boson loop is tractable, we find that at finite $T$ 
the first vertex correction is
\beq
\Gamma^{\mu\nu}_{ZZS} \sim \frac{A g^2}{4}g^{\mu\nu} \int \frac{d^3p}{(2\pi)^3} \frac{\beta \omega e^{\beta \omega}+e^{\beta \omega}-1}{\omega^3 \left(1-e^{\beta \omega}\right)^2},
\eeq 
where $\beta = 1/T$ and $\omega$ is the energy of the boson in the loop. For a massless boson, this integral is infrared divergent $\propto \int \frac{d\omega}{\omega^2}$, a well-known feature of finite-temperature corrections~\cite{Kapusta:2006pm}. However, accounting for screening with the effective thermal mass $m_{h,T}^2 = c_h T^2$, the regulated corrections scale as follows $\Gamma^{\mu\nu}_{ZZSm } \propto (Ag^2/\sqrt{c_h}) \;g^{\mu\nu}$.
Notice that despite the associated loop suppression, 
this thermal loop correction is finite in the limit $v(T)\to 0$. Therefore, near the phase transition,  a
more complete treatment of thermal loops would be necessary. 

\begin{figure*}[t]
\centering
 \includegraphics[width= 0.8\columnwidth]{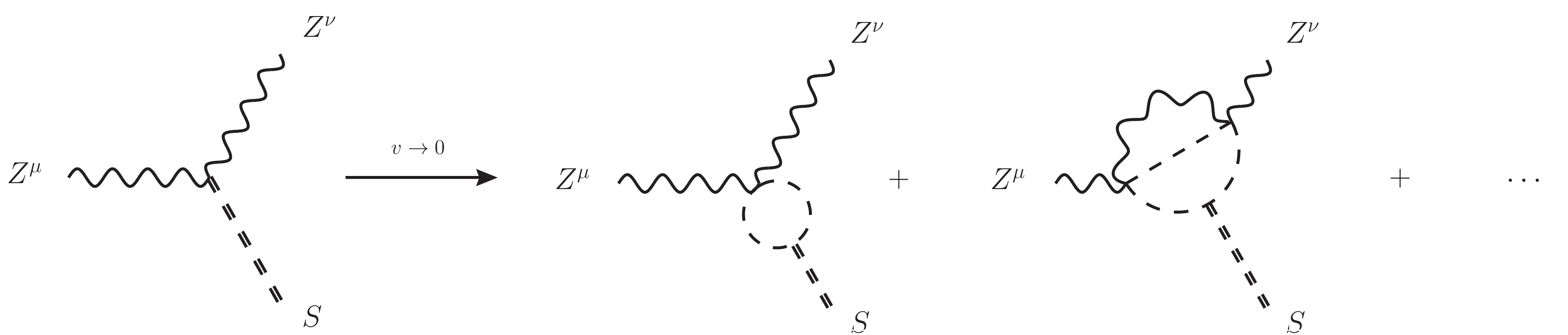}
  \caption{Survival of the ZZS vertex at higher order in the symmetric phase.}
\label{fig:ZZS_sym}
\end{figure*}

\section{$S$ production cross sections}

\label{App:Xsecs}

In this Appendix, we exhibit the production cross sections used in the calculation of the $S$ freeze-in abundance, in the limit of small $m_S$. As a couple of primary examples of QCD production, the cross sections for a gluon scattering with a top quark to produce a $S$ (in the $m_S\to 0$ limit) are
\begin{align}
\sigma_{t\bar{t}\to gS} &= \frac{\alpha_s \theta^2 y_t^2}{9 s} \left[\left(1-\frac{4m_t^2}{s}\right)\ln \left( \frac{1+\sqrt{1-\frac{4m_t^2}{s}}}{1-\sqrt{1-\frac{4m_t^2}{s}}}\right)+ \frac{8m_t^2}{s\sqrt{1-\frac{4m_t^2}{s}}}\right],
\label{eq:sigttgS}\\
\sigma_{tg\to tS} &= \frac{\alpha_s \theta^2 y_t^2}{96s}\left[ \frac{2s\left(s+3m_t^2\right)^2}{\left(s-m_t^2\right)^3}\ln\left(\frac{s}{m_t^2}\right) - \frac{1}{\left(1-\frac{s}{m_t^2}\right)^3}\left(3+ 22\frac{m_t^2}{s}-20 \frac{m_t^4}{s^2}-6 \frac{m_t^6}{s^3}+\frac{m_t^8}{s^4}\right)\right], \label{eq:sigtgtS}
\end{align}
where $y_t = \sqrt{2} m_t/v$ and $\alpha_s$ is the strong coupling constant, evaluated at the cosmic temperature $\alpha_s = \alpha_s (T)$. 

The cross sections in other channels are quite lengthy, so we collect here just the large $s$ limits to simplify the presentation. Note however that full expressions are used in the numerical computation. For the Yukawa-type annihilation, we have 
\begin{align}
\sigma_{t\bar{t}\to hS} &\to \frac{y_t^4 \theta^2}{192\pi s}\left(\ln\frac{s}{m_t^2}-2\right), \\
\sigma_{t\bar{t}\to ZS} &\to \frac{\theta^2}{576\pi v^4 s}\left[ 6m_t^2\left(2m_t^2{+}(1{+}c_v^2)m_Z^2\right) \ln \frac{s}{m_t^2} + m_Z^2\left((1+c_v^2)m_Z^2 - 24 m_t^2\right)\right], \\
\sigma_{t\bar{b}\to W^+S}&\to \frac{\theta^2}{288\pi v^4s}\left[3m_t^2(m_t^2+2m_W^2)\ln \frac{s}{m_t^2} +2m_W^2-12m_t^2m_W^2-3m_t^2\right].
\end{align}
where $c_v = I_3 - 2Q \sin^2\theta_W$, in terms of the eigenvalues of charge $Q$ and isospin $I_3$ for the relevant fermion. The leading forms for the Compton-like scattering cross sections are
\begin{align}
\sigma_{th\to tS}&\to \frac{\theta^2 y_t^4}{128\pi s}\left(2\ln\frac{s}{m_t^2}+5\right),\\
\sigma_{tZ\to tS}&\to \frac{\theta^2 m_Z^2}{48\pi v^4}(1+c_v^2) + \mathcal{O}\left(\frac{1}{s}\right), \\
\sigma_{tW^-\to b S} &\to \frac{\theta^2 m_W^2}{12\pi v^4}+ \mathcal{O}\left(\frac{1}{s}\right), \\
\sigma_{bW^+\to t S} &\to \frac{\theta^2 m_W^2}{12\pi v^4}+ \mathcal{O}\left(\frac{1}{s}\right),
\end{align}
and the bosonic scattering cross sections are
\begin{align}
\sigma_{Zh\to ZS} &\to \frac{\theta^2 m_Z^2}{12\pi v^4},\\
\sigma_{ZZ\to hS} &\to \frac{\theta^2 m_Z^2}{36 \pi v^4},\\
\sigma_{W^+W^-\to hS} &\to \frac{\theta^2 m_W^2}{18\pi v^4},  \\
\sigma_{W^+W^-\to ZS} &\to \frac{\theta^2 m_W^2 \left(8m_W^2 + m_Z^2\right)}{18\pi m_Z^2 v^4},\\
\sigma_{W^\pm h\to W^\pm S} &\to \frac{\theta^2 m_W^2}{36\pi v^4} ,\\
\sigma_{W^\pm Z\to W^\pm S} &\to \frac{\theta^2 (20m_W^4 - 3 m_W^2 m_Z^2+ m_Z^4)}{36\pi m_Z^2 v^4}, \\
\sigma_{h h\to h S} &\to \frac{9 \theta^2 \lambda^2}{8\pi s}.
\end{align}

\section{Numerical integration with quantum statistics}

\label{sec:sYdot-fullstats}

We use an integration strategy based on Ref.~\cite{Hannestad:1995rs} in which the authors reduced the collision integral in the context of neutrino decoupling from 9D to 2D retaining the quantum distributions of particles. The heavy mediator limit was assumed in this reference, which is not appropriate here, but we present a strategy to reduce the number of integrals requiring numerical treatment.

We wish to integrate
\begin{align}
s\dot{Y} &= \int \prod_{i=1}^4 \left(\frac{d^3\nvec{p}_i}{2E_i(2\pi)^3}\right) \Lambda (f_1,f_2,f_3,f_4) \times |\mathcal{M}|^2 (2\pi)^4 \delta^4(p_1+p_2-p_3-p_4),
\end{align}
where $\Lambda$ represents the thermal distribution of each species and $|\mathcal{M}|^2$ is the spin-summed squared amplitude. Working in a reference frame where species 1 travels in the $\hat{x}$ direction, we define the four-vectors
\begin{align}
p_1 &= (E_1,\nvec{p}_1,0,0), \\
p_2 &= (E_2,\nvec{p}_2 \cos\alpha,\nvec{p}_2 \sin\alpha \sin\beta,\nvec{p}_2 \sin\alpha \cos\beta) ,\\
p_2 &= (E_3,\nvec{p}_3 \cos\theta,0,\nvec{p}_3 \sin\theta) ,\\
p_4 &= p_1+p_2-p_3,
\end{align}
where $\nvec{p_i}= |\vec{p}_i|$; the angle between $\vec{p}_1$ and
$\vec{p}_2$ is $\alpha$ and between $\vec{p}_1$ and $\vec{p}_3$ is
$\theta$. Both $\vec{p}_2$ and $\vec{p}_3$ have an azimuthal angle
with $\vec{p}_1$, but there is an overall azimuthal symmetry and only
the difference between the 2 azimuthal angles matters, denoted by
$\mu$. We have used the azimuthal symmetry to fix the $\vec{p}_3$
azimuthal angle to 0. Then we have
$d^3\nvec{p}_1 d^3\nvec{p}_2 d^3\nvec{p}_3 = \nvec{p}_1 E_1
dE_1d\Omega_1 \;\nvec{p}_2 E_2 dE_2d(\cos \alpha) d\beta \;\nvec{p}_3
E_3 dE_3 \;d(\cos \theta)\; d\mu$ and our overall integral reduces to
\begin{align}
s\dot{Y} &= \frac{2(2\pi)^2}{8(2\pi)^{8}}\int \prod_{i=1}^3 \left(\nvec{p}_idE_i\right)\frac{ d^3p_4}{2E_4} d(\cos \alpha)\; d\beta \;d(\cos \theta)\, \Lambda (f_1,f_2,f_3,f_4) |\mathcal{M}|^2 \delta^4(p_1+p_2-p_3-p_4),
\end{align}
on performing the trivial integrals over $\Omega_1$ and $\mu$. We recall that the 3-dimensional integral $d^3\nvec{p}_4$ comes from $\frac{d^3\nvec{p}_4}{2E_4} = d^4p_4 \delta (p_4^2 - m_4^2) \Theta (p_4^0)$,
and we can use the 4D $\delta$-function to perform the $d^4p_4$ integral,
\begin{align}
s\dot{Y} &= \frac{1}{4(2\pi)^{6}}\int \prod_{i=1}^3 \left(\nvec{p}_idE_i\right)\ d(\cos \alpha)\; d\beta \;d(\cos \theta)\,\Lambda (f_1,f_2,f_3,f_4) |\mathcal{M}|^2  \delta ( p_4^2 - m_4^2) \Theta (p_4^0),
\end{align}
which fixes $p_4^2 = p_1^2 + p_2^2 + p_3^2 + 2 (p_1\cdot p_2 - p_1\cdot p_3 - p_2 \cdot p_3)$ from now on. The dot products can be evaluated via our angle definitions ($p_i \cdot p_j \equiv p_{ij}$)
\begin{align}
p_{12} &= E_1 E_2 - \nvec{p}_1 \nvec{p}_2 \cos \alpha,  \\
 p_{23}  &= E_2 E_3 - \nvec{p_2}\nvec{p_3}(\cos \alpha \cos \theta + \sin \alpha \sin \theta \cos \beta), \nonumber \\ 
p_{13}  &= E_1 E_3 - \nvec{p}_1 \nvec{p}_3 \cos \theta, \\
 p_{24}  &= m_2^3 +  p_{12}  -p_{23} , \\ 
p_{14} &= m_1^2 + p_{12}  -p_{13} ,  \\
 p_{34}&=  -m_3^3 +  p_{13} +p_{23} . 
\end{align}
The argument of the last delta function can be expressed as a function of $\beta$
\begin{align}
f(\beta) &= p_4^2 - m_4^2 \\
&= \omega + 2\left( \nvec{p}_2 \nvec{p}_3 \cos \alpha \cos \theta + \nvec{p}_2 \nvec{p}_3 \sin \alpha \sin \theta \cos \beta - \nvec{p}_1 \nvec{p}_2 \cos \alpha  \right), \nonumber 
\end{align}
where 
\begin{align}
\omega &= Q + 2 \left(E_1 E_2 - E_1 E_3 - E_2 E_3 + \nvec{p}_1 \nvec{p}_3 \cos \theta \right)\nonumber
\end{align}
with $Q{=}m_1^2{+}m_2^2{+}m_3^2{-}m_4^2$.
The $\beta$ integral can evaluated using $f'(\beta){=}-2\nvec{p}_2\nvec{p}_3\sin\alpha \sin \theta \sin \beta$, forcing $\beta \to \beta_0$, where
\beq
\cos \beta_0 = -\frac{ \omega +2\left( \nvec{p}_2 \nvec{p}_3 \cos \alpha \cos \theta - \nvec{p}_1 \nvec{p}_2 \cos \alpha  \right) }{2\nvec{p}_2 \nvec{p}_3 \sin \alpha \sin \theta}
\eeq
is found by solving $f(\beta_0) = 0$. There are actually two $\beta_0$ solutions given by $\sin \beta_0 = \pm \sqrt{1-\cos^2 \beta_0}$. Since everything is symmetric in $\beta$ (all dot products are $\cos \beta$-dependent and the $\partial f/\partial \beta$ factor that appears in the dominator is an absolute value), we can simply use the positive root and multiply by 2. Hence, we get
\beq
s\dot{Y} = \frac{1}{2(2\pi)^{6}}\int \prod_{i=1}^3 \left(\nvec{p}_idE_i\right)\ d(\cos \alpha) \;d(\cos \theta)\; \Lambda (f_1,f_2,f_3,f_4)  \frac{|\mathcal{M}|^2  \Theta (p_4^0)\Theta(4\nvec{p}^2_2 \nvec{p}^2_3 \sin^2 \alpha \sin^2 \theta \sin^2 \beta_0)}{2\nvec{p}_2 \nvec{p}_3 \sin \alpha \sin \theta \sin \beta_0}.
\eeq
The extra step-function arises via an obligation to maintain $\beta_0$ in the physical phase-space
\begin{align}
\cos^2 \beta_0 \leq 0 \quad  &\leftrightarrow \quad(2\nvec{p}_2 \nvec{p}_3 \sin \alpha \sin \theta \sin \beta_0)^2 \geq 0 \nonumber \\ 
&\leftrightarrow \quad \left|\frac{\partial f}{\partial \beta_0} \right|^2 \geq 0.
\end{align}
We can now focus on the angular integrations.
\begin{align}
s\dot{Y} &= \frac{1}{2(2\pi)^{6}}\int dE_1dE_2dE_3 \; \nvec{p}_1 \nvec{p}_2 \nvec{p}_3 \Lambda (f_1,f_2,f_3,f_4) \times \mathcal{I} \nonumber\\
\mathcal{I} &= \int  d(\cos \theta) d(\cos \alpha)\;\frac{|\mathcal{M}|^2  \Theta (p_4^0)\Theta\left(\left|\frac{\partial f}{\partial \beta_0} \right|^2\right )}{\left|\frac{\partial f}{\partial \beta_0} \right|}.
\end{align}
Expanding $f'$ as 
\begin{align}
\left|\frac{\partial f}{\partial \beta_0} \right|^2 &= a \cos^2 \alpha + b\cos \alpha + c \\
a &= -4\nvec{p}_2^2 \left(\nvec{p}_1^2 + \nvec{p}_3^2 - 2\nvec{p}_1\nvec{p}_3 \cos \theta\right)\\
b &= 4 \nvec{p}_2 \left(\nvec{p}_1 - \nvec{p}_3 \cos\theta\right) \omega\\
c &= 4\nvec{p}_2^2 \nvec{p}_3^2 \sin^2 \theta - \omega^2,
\end{align}
the step-function ensures that the denominator is real,
\beq
\mathcal{I} = \int  d(\cos \theta) \int_{\cos \alpha_-}^{\cos \alpha_+}d(\cos \alpha)\;\frac{|\mathcal{M}|^2  \Theta (p_4^0)\Theta\left(a\cos^2\alpha + b\cos \alpha + c\right )}{\sqrt{a\cos^2\alpha + b\cos \alpha + c}}.
\eeq
Since $|\mathcal{M}|^2$ only consists of simple functions of $\cos \alpha$, the $\cos \alpha$ integration can be performed straightforwardly for each process. Since $a\leq 0$, the integration bounds are set by the real-valued criterion, between which the quadratic function is positive. Given the roots $\cos \alpha_\pm =  (-b\mp \sqrt{b^2 - 4ac})/(2a)$, 
notice that we always have $-1 \leq \cos \alpha_- $ and $\cos \alpha_+ \leq 1$. The $\theta$ integral can be performed in a similar fashion. Requiring $\cos \alpha_\pm$ to be real implies the condition $b^2-4 a c \geq 0$, and
\begin{align} 
&\cos \theta_{\pm} = -\frac{Q + 2\nvec{p}_2^2 + 2 \gamma \mp 2 \nvec{p}_2 \sqrt{Q+\nvec{p}_1^2 + \nvec{p}_2^2 + \nvec{p}_3^2 + 2\gamma}}{2\nvec{p}_1\nvec{p}_3}, \nonumber
\end{align}
where we use the shorthand $\gamma = E_1E_2 - E_1 E_3 - E_2 E_3$ and we have
\beq
\mathcal{I} = \int_{\max(-1,\cos\theta_-)}^{\min(1,\cos\theta_+)}  d(\cos \theta) \int_{\cos \alpha_-}^{\cos \alpha_+}d(\cos \alpha)\;\frac{|\mathcal{M}|^2  \Theta (p_4^0)\Theta\left(a\cos^2\alpha + b\cos \alpha + c\right )}{\sqrt{a\cos^2\alpha + b\cos \alpha + c}}.
\eeq
These 2 integrals can be carried out analytically for each reaction. The final integral to be performed numerically for the emissivity is 3-dimensional, and a step function guarantees that the phase-space is physical and $\cos \alpha_\pm$ is real-valued,
\begin{align}
s\dot{Y} &= \frac{1}{2(2\pi)^{6}}\int dE_1dE_2dE_3 \; \nvec{p}_1 \nvec{p}_2 \nvec{p}_3 \Lambda (f_1,f_2,f_3,f_4) \times \mathcal{I} \times \Theta (Q+ \nvec{p}_1^2 + \nvec{p}_2^2 + \nvec{p}_3^2 + 2\gamma) \\
\mathcal{I} &= \int_{\max(-1,\cos\theta_-)}^{\min(1,\cos\theta_+)}  d(\cos \theta) \int_{\cos \alpha_-}^{\cos \alpha_+}d(\cos \alpha)\;\frac{|\mathcal{M}|^2}{\sqrt{a\cos^2\alpha + b\cos \alpha + c}}.
\end{align}

\end{widetext}

\end{document}